\documentclass[twocolumn,aps,prd,10pt,showpacs,preprintnumbers,nofootinbib,superscriptaddress]{revtex4-1}
\usepackage{graphicx,slashed,bbold,subfigure,amsmath,amssymb,hyperref,enumerate}
\usepackage[english]{babel}

\newcommand{\dt}{{\mathrm{det}}}

\newcommand{\ud}{\mathrm{d}}
\newcommand{\st}{{\mathrm{st}}}

\newcommand{\eff}{\mathrm{eff}}

\begin{document}
\title{Thermodynamical properties of strongly interacting matter in a model with explicit chiral symmetry breaking interactions} 
\author{J. Moreira}
\affiliation{CFisUC, Department of Physics, University of Coimbra, P-3004-516 Coimbra, Portugal}
\email{jmoreira@uc.pt}
\author{J. Morais}
\affiliation{CFisUC, Department of Physics, University of Coimbra, P-3004-516 Coimbra, Portugal}
\email{jorge.m.r.morais@gmail.com}
\author{B. Hiller}
\affiliation{CFisUC, Department of Physics, University of Coimbra, P-3004-516 Coimbra, Portugal}
\email{brigitte@teor.fis.uc.pt}
\author{A. A. Osipov}
\affiliation{Bogoliubov Laboratory of Theoretical Physics, JINR, Dubna, Moscow Region,Russia}
\email{aaosipov@jinr.ru}
\author{A. H. Blin}
\affiliation{CFisUC, Department of Physics, University of Coimbra, P-3004-516 Coimbra, Portugal}
\email{alex@teor.fis.uc.pt}

\begin{abstract}
We analyse the effects of the light and strange  current quark masses  on the phase diagram of QCD at finite temperature and vanishing baryonic chemical potential, computing the speed of sound, the trace anomaly of the energy  momentum
tensor and the fluctuations and correlations of the conserved charges associated to baryonic, electric and strangeness numbers. The framework is a known extension of the three flavor Nambu Jona Lasinio model, which includes the full set of explicit chiral symmetry breaking interactions (ESB) up to the same order in large $N_c$ counting as the 't Hooft flavor mixing terms and eight quark interactions. It is shown that the ESB terms are relevant for the description of a soft region in the system's speed of sound and overall slope behavior of the observables computed. At the same time the role of the 8q interactions gets highlighted. The model extension with the Polyakov loop is considered and the results are compared to lattice QCD data.  
\end{abstract}

%\begin{keyword}
%explicit chiral symmetry breaking \sep Nambu--Jona-Lasinio model \sep phase diagram \sep strange quark matter
%\end{keyword}
%,14.40.Aq,12.40.Yx}
\maketitle

\section{Introduction}
The study of the thermodynamical properties of strongly interacting matter is one of the open present day theoretical and experimental challenges. On the theoretical side the well known problems of the \textit{ab initio} approach of lattice QCD (lQCD) when dealing with finite chemical potential as well as the will to grasp a deeper understanding of the interplay of the various underlying mechanisms in strongly interacting matter can be seen as an incentive to the use of moderately complex effective lagrangians. On the other hand at vanishing chemical potential the growing confidence in lQCD results means that the agreement with these is increasingly used as a way to establish the success of other theoretical predictions. 

The values of the current quark masses constitute undoubtedly one of the most relevant inputs in the study of the QCD phase diagram, as the quark condensates become not exact order parameters for the chiral phase transition. %The concept of $restoration of chiral symmetry at finite baryonic density and temperature gets blurred,
Model estimates of the size of the condensates at the critical transition points show that they may be significantly larger than the bare ones \cite{Asakawa:1989bq,Klevansky:1992qe,Ebert:1992ag,Hatsuda:1994pi,Buballa:2003qv,Osipov:2007mk,Fukushima:2013rx}, indicating that non-perturbative effects are still effective in spite of the transition. After the transition, a more or less slow convergence to the bare values of the condensates depends naturally on the size of the current quark masses and how fast the perturbative regime of QCD is reached, where bulk thermodynamic observables are conditioned by the Stefan-Boltzmann limit pertinent to an ideal quark-gluon gas. 

The chiral critical end point (CEP) which according to numerous model calculations is expected to occur, separating a region of first order transitions at higher baryon chemical potential $\mu_B$ and lower temperatures T from the crossover behavior at lower $\mu_B$ and higher T, is not yet established.  A second order transition is likely to occur at this point in the chiral limit of light quarks  and an infinitely heavy strange quark  \cite{Rajagopal:1999cp}.  Model results, using quark masses close to physical values, differ drastically regarding its possible location. The hope that its eventual location may be narrowed down in  lQCD, using the reweighting technique to extend lattice calculations from $\mu_B=0$ to finite values \cite{Fodor:2001pe}, should the nature of the transition be second order,  requires still a detailed analysis of the quark mass and volume dependence \cite{Ejiri:2005ts}, on which hinges the accuracy of Lee-Yang zeros of the lQCD partition function \cite{Karsch:2006xs}. 
 
Meanwhile extensive lQCD studies by different groups report that the algorithmic difficulties that prevented the use of the light physical
quark masses have been mostly overcome, as well as spurious taste breaking effects for staggered discretization schemes, making it possible to achieve a realistic hadron spectrum \cite{Aoki:2012oma,Bazavov:2011nk}. 
Their findings converge to the by now commonly accepted understanding that along the $\mu_B=0$ line no genuine phase transition occurs. A  crossover takes place around $T\sim 155$ MeV  in a T interval of roughly 20 MeV \cite{Borsanyi:2013bia}, \cite{Bazavov:2014pvz},  for   recent reports  see \cite{Borsanyi:2016bzg},
 \cite{universe3010007}
.  
This value of T decreased substantially as compared to the quoted value one decade ago, $T=192$ MeV \cite{Cheng:2006qk}, that used calculations with
improved staggered fermions for various light to strange quark mass ratios in the range $ [0.05, 0.5]$,  and with a strange quark mass fixed close to its physical value (although with an estimate for the string tension
10$\%$ larger than the usually quoted), while in \cite{Aoki:2006br} the crossover temperature was reported to be close to the present day value for the renormalized chiral susceptibility and about 25 MeV larger for the strange
quark number susceptibility and Polyakov loops,  using physical quark mass values. For comparison, the RHIC freeze-out values  indicated at that time occur below $T\sim 170$ MeV \cite{Adcox:2004mh,Adams:2005dq}. 

The necessity to devise powerful measures of signatures for this crossover,  that could also be useful in the experimental searches,  has become a main objective. Besides  the chiral condensate and its derivative with respect to the quark mass, the chiral susceptibility, used  to  probe  the  restoration  of  chiral  symmetry,  fluctuations and correlations in conserved charges have become  tools to identify the transition from hadronic to quark-gluon degrees of freedom in the crossover region. In relativistic heavy-ion colliders experiments the ratios of such fluctuations are obtained in precision experimental studies for several collision energies as part of the RHIC beam energy scan program \cite{Adamczyk:2013dal,Adamczyk:2013caa}, which focuses on the search of the CEP.  Chemical freeze-out parameters are then 
extracted within a canonical ensemble description of the data. These parameters lie below the freeze-out parameters extracted formerly from particle yields in the Hadron Resonance Gas (HRG) model  \cite{Cleymans:1998yb} ,\cite{BraunMunzinger:2001ip,BraunMunzinger:2003zd} and the single freeze-out model \cite{Broniowski:2001we,PhysRevC.65.064905,Broniowski:2003ax}.
A recent analysis using the HRG model to fit the net Kaon fluctuations at RHIC \cite{Bellwied:2018tkc}  provides experimental evidence  that  the freeze-out temperatures for strange hadrons could be  $10 - 15$  MeV  higher than for the light ones. 

Other measures resort to the equation of state (EoS) for determination of the trace of the energy momentum tensor and related specific heat and speed of sound. It has been pointed out a long time ago  that the EoS near the QCD phase transition might be very soft as compared to an ideal pion gas \cite{Shuryak:1986nk,VanHove:1983rk, Kajantie:1986dh} such that a "longest lived fireball" could be produced in relativistic heavy ion collisions \cite{Hung:1994eq} with ideal conditions to study signatures of the quark gluon plasma (QGP) as it goes through the stage of deconfinement. This softest point  in the EoS is seen as a minimum in the velocity of sound, which measures the rate of change of the pressure with respect to the energy density.  lQCD results show that at the minimum the energy density is only slightly above that of normal nuclear matter density \cite{Bazavov:2014pvz}.

In the present study we address the effects of the light and strange current quark masses in the light of an effective theory of QCD. As is well known from Chiral Perturbation Theory the canonical mass term represents only the leading order of an expansion in the masses themselves \cite{Weinberg:1978kz,Gasser:1982ap,Gasser:1984gg}. The explicit symmetry breaking pattern involves  current quark mass dependent interactions at higher orders. 

We use a three flavor Nambu-Jona-Lasinio \cite{Nambu:1961tp,Nambu:1961fr} related Lagrangian in mean field approximation, which we have previously extended to include relevant spin zero interactions up to the same order in large $N_c$ counting as the $U(1)_A$ breaking flavor determinantal interaction of 't Hooft  \cite{'tHooft:1976fv, PhysRevD.18.2199.3}.  In the context of the NJL model the 't Hooft term has been first introduced in \cite{Kunihiro:1987bb,Bernard:1987sg,Reinhardt:1988xu}; in this form the model has been extensively used, see e.g. \cite{Vogl:1989ea,Klimt:1989pm,Takizawa:1990ku,Klevansky:1992qe,Bernard:1993wf,Hatsuda:1994pi,Birse:1996dx,Dmitrasinovic:2000ei,Naito:2003zj,Buballa:2003qv,Volkov:2005kw,Fukushima:2013rx}. 

Our extension  included first two kinds of eight quark  chiral symmetry preserving interactions \cite{Osipov:2005tq,Osipov:2006ns}, which were needed to complete the number of vertices important for dynamical chiral symmetry breaking  in four dimensions \cite{Andrianov:1993wn, Andrianov:1992tb}, and resolved instability issues related to the model's effective potential reported in \cite{Osipov:2005sp}. The next extension added the set of explicit symmetry breaking (ESB) multiquark interactions at the specific $N_c$ order considered \cite{Osipov:2012kk,Osipov:2013fka}. The phenomenological impact of the ESB terms on the quality of the low lying spectra of pseudoscalars and scalars as well as other related observables is remarkable, in comparison with the models without their inclusion. In particular the possibility to accurately describe the spectra of the scalar mesons together with a good fit for their strong decays gave us confidence that the model parameters obtained represented an adequate set for the analysis of the model related QCD  phase diagram.

In a subsequent work   \cite{Moreira:2014qna} we have shown that the extended model leads to the emergence of two CEPs, associated with the light and strange quark condensates, in contrast to the  common picture in which only the light condensate relates to a first order transition (except for a small effect on the strange condensate due to the coupling of both sectors), while the strange condensate displays a crossover behavior. The two CEPs act upon the onset for formation of strange quark matter, which is shifted  to significantly smaller values of $\mu_B$.

Recently a similar extension of model interactions within the quark meson model and taking into account finite size effects led to the interesting result that quark matter with only u and d quarks may be the stable configuration in a region close to the end of the table of elements \cite{Holdom:2017gdc}.

In face of these new developments one recognizes that the current quark mass effects are far from being fully explored and understood.

The crossover regime which follows after the two CEP's reported in \cite{Moreira:2014qna}  towards lower $\mu_B$ until reaching the $\mu_B=0$ line at higher temperatures is expected to weaken the prominent ESB features mentioned for the critical zone. Nevertheless we show that some observables are sensitive to the current quark mass values and ESB interactions. We stress that the numerical values of the current quark masses are intertwined with the dynamics of the ESB interaction terms and must always be considered together in the extended version.  The numerical value  of the strange current quark mass is reduced by roughly a factor 2 (thus being possible to reach its empirical value and light to strange quark mass ratios) when ESB interactions are considered in conjunction with best fits for the hadronic spectra, compared to the case without ESB interactions. 

The obvious drawback of the model is the lack of confinement. However, it should be noted that the NJL model shares the global symmetries with QCD. Therefore it gives a reasonable tool to study the critical phenomena even if the location of the CEP may differ. For instance, both NJL and Polyakov-loop NJL models predict that the critical point is located inside the pion condensed phase. This result is consistent with the QCD no-go theorem \cite{Hidaka:2011jj}, which is  a rigorous statement in the large-$N_c$ limit.  Thus, we suppose that symmetries combined with the $1/N_c$ approach lead to a reasonable picture of the phase properties of the hadron matter even if the confinement mechanism is not included. 
 
The comparison with lQCD data shows that there is room to improve the model calculations, mainly with respect to the transition to the Stefan-Boltzmann limit that occurs in a much shorter temperature interval, as well as a  systematic shift to lower temperatures of the observables considered. The main observation is that the model is able to capture important features, such as the presence of a dip in the sound velocity, and  the slopes of most observables considered. 

Polyakov loop extended versions of NJL model have been extensively used in the study of temperature effects in strongly interacting matter  \cite{Fukushima:2003fw, Megias:2003ui,Megias:2004hj,Roessner:2006xn,Ghosh:2007wy,Fu:2007xc,Costa:2008dp,Fukushima:2008wg,Bhattacharyya:2010wp,Moreira:2010bx,Stiele:2016cfs,Bhattacharyya:2016jsn}. 

Here we show that by coupling the model to the gluonic degrees of freedom through the Polyakov loop, the temperature gap between our model predictions and lQCD data is practically removed, as well as an improvement is obtained regarding the height of the peak of one of the correlators. An overall good agreement with lQCD data is  obtained.

The text is organized as follows: after revisiting the model Lagrangian and thermodynamic potential in section   \ref{Model} we present the thermodynamic observables that we compute and the model fits in section \ref{therm},\ref{pol} and \ref{fits} and discuss the results in section \ref{Results}. Our conclusions are summarized in section   \ref{Conclusions}.

%%%%%%%%%%%%%%%%%%%%%%%%%%%%%
%%%%%%%%%%%%%%%%%%%%%%%%%%%%%

\section{The Model}
\label{Model}
\subsection{Model thermodynamic potential}
Although the model Lagrangian has been introduced and applied in previous works we indicate it here for completeness and refer for further details to \cite{Osipov:2012kk,Osipov:2013fka,Moreira:2014qna,Osipov:2015lva}.

The effective multiquark Lagrangian is expressed in terms of the $U(3)$ Lie-algebra valued field $\Sigma =(s_a-ip_a)
\frac{1}{2}\lambda_a$, involving the quark bilinears $s_a=\bar q\lambda_aq$, $p_a=\bar q\lambda_ai
\gamma_5q$;  $a=0,1,\ldots ,8$, $\lambda_0=\sqrt{2/3}\times 1$, $\lambda_a$ 
are the standard $SU(3)$ Gell-Mann matrices for $1\leq a \leq 8$. The $q$ designates the color quark fields which enjoy the chiral flavor $U(3)_L\times U(3)_R$ global symmetry of QCD in the massless case. 
In addition the Lagrangian depends on external sources 
$\chi$, which generate explicit symmetry breaking effects. In terms of these fields and sources the Lagrangian density reads to next to leading order (NLO) in the large $N_c$ counting
 \begin{equation}
\label{LQ}
   {\cal L}_\eff=\bar qi\gamma^\mu\partial_\mu q+{\cal L}_{int}+{\cal L}_\chi,
\end{equation}
with
\label{ModelPotential}
\begin{align}
\label{L-int}
{\cal L}_{\mathrm{int}}&=\frac{\bar G}{\Lambda^2}\mbox{tr}\left(\Sigma^\dagger\Sigma\right)+\frac{\bar\kappa}{\Lambda^5}\left(\det\Sigma+\det\Sigma^\dagger\right) \nonumber \\
&+\frac{\bar g_1}{\Lambda^8}\left(\mbox{tr}\,\Sigma^\dagger\Sigma\right)^2+\frac{\bar g_2}{\Lambda^8}\mbox{tr}\left(\Sigma^\dagger\Sigma\Sigma^\dagger\Sigma\right).
\end{align}  
and the source dependent pieces 
\begin{align}
  {\cal  L}_\chi &=\sum_{i=0}^{10}L_i, \nonumber \\
   L_0&=-\mbox{tr}\left(\Sigma^\dagger\chi +\chi^\dagger\Sigma\right), \nonumber \\
   L_1&=-\frac{\bar\kappa_1}{\Lambda}e_{ijk}e_{mnl}\Sigma_{im}\chi_{jn}\chi_{kl}+h.c.\nonumber \\
   L_2&=\frac{\bar\kappa_2}{\Lambda^3}e_{ijk}e_{mnl}\chi_{im}\Sigma_{jn}\Sigma_{kl}+h.c.,\nonumber \\
   L_3&=\frac{\bar g_3}{\Lambda^6}\mbox{tr}\left(\Sigma^\dagger\Sigma\Sigma^\dagger\chi\right)+h.c.\nonumber \\
   L_4&=\frac{\bar g_4}{\Lambda^6}\mbox{tr}\left(\Sigma^\dagger\Sigma\right)\mbox{tr}\left(\Sigma^\dagger\chi\right)+h.c.,\nonumber \\
   L_5&=\frac{\bar g_5}{\Lambda^4}\mbox{tr}\left(\Sigma^\dagger\chi\Sigma^\dagger\chi\right)+h.c.\nonumber \\
   L_6&=\frac{\bar g_6}{\Lambda^4}\mbox{tr}\left(\Sigma\Sigma^\dagger\chi\chi^\dagger +\Sigma^\dagger\Sigma\chi^\dagger\chi\right),\nonumber \\
   L_7&=\frac{\bar g_7}{\Lambda^4}\left(\mbox{tr}\Sigma^\dagger\chi+ h.c.\right)^2\nonumber \\ 
   L_8&=\frac{\bar g_8}{\Lambda^4}\left(\mbox{tr}\Sigma^\dagger\chi- h.c.\right)^2,\nonumber \\ 
   L_9&=-\frac{\bar g_9}{\Lambda^2}\mbox{tr}\left(\Sigma^\dagger\chi\chi^\dagger\chi\right)+h.c.\nonumber \\
   L_{10}&=-\frac{\bar g_{10}}{\Lambda^2}\mbox{tr}\left(\chi^\dagger\chi\right)\mbox{tr}\left(\chi^\dagger\Sigma\right)+h.c.
\end{align}
Under
chiral transformations one has for the quark fields $q'=V_Rq_R+V_Lq_L$, where $q_R=P_R q, q_L=P_Lq$, and 
$P_{R,L}=\frac{1}{2}(1\pm\gamma_5)$. Then $\Sigma'=V_R\Sigma V_L^\dagger$, and 
$\Sigma^{\dagger'}=V_L\Sigma^\dagger V_R^\dagger$; the sources transform as the field $\Sigma$.

At this stage  the sources can be fixed as the current quark masses $\chi=1/2 \mbox{diag}(m_u,m_d,m_s)$, after using the freedom related with the Kaplan-Manohar ambiguity
associated with $L_1,L_9,L_{10}$ \cite{Kaplan:1986ru} and which are henceforth set to zero.

The couplings $g_i={\bar g_i}/\Lambda^n$ ($g_i$ stands generically for any coupling) carry negative dimensions, given by the powers of $\Lambda$, and thus the Lagrangian is non renormalizable. 
% all barred coulings ${\bar G},{\bar\kappa},{\bar g_1},{\bar g_2},{\bar\kappa_1},{\bar\kappa_2},{\bar g_3}...{\bar g_{10}$ are dimensionless,  
 Here $\Lambda\sim 4 \pi f_\pi \sim 1$ GeV \cite{Manohar:1983md} is associated with the scale of spontaneous chiral symmetry breaking.  

As mentioned in the introduction this Lagrangian contains all non-derivative spin 0 multiquark interactions up to the same counting in large $N_c$ as the 't Hooft interaction, 
given by the term $\sim\kappa$ in Eq. \ref{L-int}. It has been shown that the large $N_c$ counting scheme selects the same interactions that are also relevant in the effective
 potential in the limit $\Lambda \rightarrow\infty$, i.e. those  scaling at most as $\Lambda^0$. The LO contributions are only the $4q$ interaction $\sim G$ of the original NJL
 Lagrangian and the canonical quark mass term $L_0$, all the other terms are $N_c^{-1}$ suppressed with respect to LO. These contain terms which violate the OZI 
rule  $(\kappa,\kappa_2, g_1, g_4, g_7, g_8)$, of which $(\kappa,\kappa_2)$ break the $U(1)_A$ symmetry and are thus anomalous, as well as interactions which describe 
 four-quark component  $\bar qq\bar qq$ admixtures to the $\bar qq$ ones ($g_2, g_3, g_5, g_6$). 

The bosonization of the Lagrangian is carried over with functional integral techniques in the stationary phase approximation resulting in the 
ef\mbox{}fective 
mesonic Lagrangian  density ${\cal L}_{{\rm bos}}$ at $T=\mu=0$,
\vspace{0.5cm}
\begin{eqnarray}
\label{bos}
    &&{\cal L}_\eff\to {\cal L}_{{\rm bos}}={\cal L}_{\st}+{\cal L}_{{\rm ql}}, 
\nonumber\\
 %   \qquad
   &&{\cal L}_{\st}=h_a\sigma_a+\frac{1}{2} h_{ab}^{(1)}
                  \sigma_a\sigma_b+\frac{1}{2} h_{ab}^{(2)}
                  \phi_a\phi_b +{\cal O}({\mbox{f\mbox{}ield}}^3),
                  \nonumber\\
    &&W_{{\rm ql}}(\sigma,\phi )=\frac{1}{2}\mbox{ln}|\dt 
                 D^{\dagger}_E D_E|=-\!\int\!
                 \frac{\ud^4 x_E}{32 \pi^2}
                 \sum_{i=0}^\infty I_{i-1}\mbox{tr}(b_i),
                 \nonumber\\
    &&b_0 =1,\hspace{0.2cm} b_1=-Y, \nonumber \\
    &&              b_2=\frac{Y^2}{2} 
                 +\frac{\Delta_{ud}}{2}\lambda_3 Y    +\frac{\Delta_{us}+\Delta_{ds}}{2\sqrt{3}}\lambda_8 
                 Y, \hspace{0.2cm} \ldots , 
                 \nonumber\\
    &&Y=i\gamma_{\alpha}(\partial_{\alpha}\sigma 
                 +i\gamma_5\partial_{\alpha}\phi )
                 +\sigma^2+\{{\cal M},\sigma\} \nonumber \\
&& \hspace{0.5cm}  +\phi^2 +i\gamma_5[\sigma+{\cal M},\phi ],  
\end{eqnarray}
written in terms of the scalar, $\sigma=\lambda_a\sigma_a$, and pseudoscalar, 
$\phi=\lambda_a\phi_a$, nonet valued f\mbox{}ields. 

The result of the 
stationary phase integration at leading order, ${\cal L}_\st$, is shown here 
as a series in growing powers of $\sigma$ and $\phi$. The coefficients $h_a,h_{ab},...$ depend
on the current quark masses and encode all the dependence in the coupling constants, see Eq. \ref{spc} below for $h_a$. As in the case of the mass parameters, 
also only the $h_a$ with $(a=0,3,8)$  (or $h_i$, $(i=u,d,s)$ in the flavor basis) do not vanish \cite{Osipov:2006ns}.

 The result of the remaining
Gaussian integration over the quark f\mbox{}ields is given by 
$W_{\rm ql}$. Here the second order operator in euclidean space-time ${D^{\dagger}_E D}_E =
{\cal M}^2-\partial_\alpha^2+Y$ is associated with the euclidean Dirac operator 
$D_E=i\gamma_\alpha\partial_\alpha -{\cal M}-\sigma -i\gamma_5\phi$ (the 
$\gamma_\alpha, \alpha =1,2,3,4$ are antihermitian and obey $\{\gamma_\alpha,
\gamma_\beta\}=-2\delta_{\alpha\beta}$); ${\cal M}=\mbox{diag}(M_u,M_d,M_s)$ is 
the constituent quark mass matrix resulting from the process of spontaneous symmetry breaking, which requires a redefinition of the field
$\sigma\rightarrow\sigma+{\cal M}$, such that the new vacuum expectation value vanishes $<\sigma>=0$.
The one-quark-loop action $W_{\rm ql}$ has been obtained 
by using a modif\mbox{}ied inverse mass expansion of the heat kernel 
associated with the given second order operator \cite{Osipov:2001th,Osipov:2001bj}. The procedure takes into 
account the dif\mbox{}ferences $\Delta_{ij}=M_i^2-M_j^2$, in a chiral invariant way at each order of 
the expansion, with $b_i$ being the generalized Seeley--DeWitt 
coef\mbox{}f\mbox{}icients.
The
\begin{equation}
\label{I-i}
   I_i=\frac{1}{3}\left[J_i(M_u^2)+J_i(M_d^2)+J_i(M_s^2)\right]
\end{equation}
is the average over the regularized 1-loop euclidean momentum integrals $J_i$ with $i+1$ vertices 
($i=0,1,\ldots$)
\begin{eqnarray} 
\label{Ji}
  &&  J_i(M^2)=16\pi^2\Gamma (i+1)\!\int %_{{\mathrm R}}
             \frac{\ud^4p_E}{(2\pi)^4}\,\hat\rho_\Lambda 
             \frac{1}{(p_E^2+M^2)^{i+1}} \nonumber \\
&&= 16\pi^2  \int \frac{\ud^4p_E}{(2\pi)^4}\,\int_0^{\infty} d\tau \hat\rho_\Lambda  \tau^i e^{-\tau (p_E^2+M^2)}. 
\end{eqnarray}  
We use the Pauli--Villars 
regularization \cite{Pauli:1949zm} with two subtractions in the integrand \cite{Volkov:1984fr}
\begin{equation}
\label{reg-2}
    \hat\rho_\Lambda\equiv\rho(\tau \Lambda^2) =1-(1-\tau \Lambda^2)e^{-\tau \Lambda^2}.     
\end{equation} 
We take only the dominant contributions to the heat kernel series, up to $b_1,b_2$   for meson spectra and decays, which involve the logarithmically $I_1$ and quadratically $I_0$ divergent integrals in $\Lambda$. 

The $J_{-1}$ integral is obtained as \cite{Osipov:2003xu}
\begin{equation}
\label{rec}
 J_{-1}(M^2)=-\int_0^{M^2}  J_0(\alpha^2) d\alpha^2.
\end{equation}

The model thermodynamical potential $\Omega$ in the mean field approximation is written as a contribution stemming from the stationary phase approximation containing all the dependence on the model couplings, $\mathcal{V}_{st}$, and one which is related to the heat kernel quark one loop integrals $J_{-1}$ which now carry the explicit $T,\mu$ dependence (for details please see \cite{Hiller:2008nu})
\begin{align}
\label{Omega}
\Omega=&\mathcal{V}_{st}+\sum_i\frac{N_c}{8\pi^2}J_{-1}(M_i,T,\mu_i),
\end{align}

\begin{align}
J_{-1}=&J^{vac}_{-1}+J^{med}_{-1},\nonumber\\
J^{vac}_{-1}=&\int\frac{\mathrm{d}^4 p_E}{(2\pi)^4}\int^\infty_0 \frac{\mathrm{d}\tau}{\tau}\rho\left(\tau\Lambda^2\right)16\pi^2\nonumber\\
&\times\left(\mathrm{e}^{-\tau\left(p_ {0\,E}^2+\boldsymbol{p}^2+M^2\right)}-\mathrm{e}^{-\tau\left(p_ {0\,E}^2+\boldsymbol{p}^2\right)}\right),\nonumber\\
J^{med}_{-1}=&-\int\frac{\mathrm{d}^3p}{(2\pi)^3}16\pi^2T\left.\left(\mathcal{Z}^++\mathcal{Z}^-\right)\right|^{M}_{0}+C(T,\mu),\nonumber\\
\mathcal{Z}^\pm =&\mathrm{log}\left(1+\mathrm{e}^{-\frac{E\mp\mu}{T}}\right)-\mathrm{log}\left(1+\mathrm{e}^{-\frac{E_\Lambda\mp\mu}{T}}\right)\nonumber\\
&-\frac{\Lambda^2}{2T E_\Lambda}\frac{\mathrm{e}^{-\frac{E_\Lambda\mp\mu}{T}}}{1+\mathrm{e}^{-\frac{E_\Lambda\mp\mu}{T}}},\nonumber\\
C(T,\mu)=&\int\frac{\mathrm{d}^3p}{(2\pi)^3}16\pi^2T~\nonumber\\
&\times\mathrm{log}\left(\left(1+\mathrm{e}^{-\frac{|\boldsymbol{p}|-\mu}{T}}\right)\left(1+\mathrm{e}^{-\frac{|\boldsymbol{p}|+\mu}{T}}\right)\right)
\end{align}
with $E=\sqrt{M^2+p^2}$, $E_\Lambda=\sqrt{E^2+\Lambda^2}$.

\begin{align}
\label{Vst}
\mathcal{V}_{st}=&\nonumber\\
\frac{1}{16}
\bigg(&  4 G \left(h_i^2\right) + 3 g_1 \left(h_i^2\right)^2 + 3 g_2 \left(h_i^4\right) + 4 g_3 \left(h_i^3 m_i\right)\nonumber\\
			&+ 4 g_4 \left(h_i^2\right) \left(h_j m_j\right) + 2 g_5 \left(h_i^2 m_i^2\right) + 2 g_6 \left(h_i^2 m_i^2\right) \nonumber\\ 
	    &+ 4 g_7 \left(h_i m_i\right)^2 + 8 \kappa h_u h_d h_s \nonumber\\
			&+ 8 \kappa_2 \left( m_u h_d h_s + h_u m_d h_s +  h_u h_d m_s\right)\bigg)\bigg|^{M_i}_0,
\end{align}
where  $h_i$, $i=(u,d,s)$ are solutions of the following system of cubic equations  
\begin{align}
\label{spc}
\Delta_f =&  M_f-m_f\\ 
         =& -G h_f -\frac{g_1}{2} h_f(h_i^2) - \frac{g_2}{2} (h_f^3) - \frac{3 g_3}{4} h_f^2 m_f\nonumber\\
				  &	-\frac{  g_4}{4} \left( m_f \left(h_i^2\right)+2 h_f(m_i h_i)\right) - \frac{g_5 + g_6}{2} h_f m_f^2\nonumber\\
					&	- g_7 m_f (h_i m_i) - \frac{\kappa}{4} t_{fij}h_i h_j - \kappa_2 t_{fij}h_i m_j.\nonumber
\end{align}
The $h_i$ are equal to one half the (unsubtracted) quark condensates, i.e. without the 2nd term in
\begin{align}
\label{qco}
\langle \bar{q} q \rangle_i= -\frac{N_c}{4 \pi^2}(\left(M _iJ_0\left[M_i^2\right]-m_iJ_0\left[m_i^2\right]\right)).
\end{align}  

\section{Thermodynamical properties of strongly interacting matter}
\label{therm}
Strongly  interacting matter is expected to undergo two transitions when subjected to high enough temperature ($T$) and/or chemical potential ($\mu$): deconfinement and chiral symmetry (partial) restoration. Although a straight connection between the two is still unclear they are for the most part expected to occur more or less simultaneously \cite{Casher:1979vw}, \cite{Banks:1979yr}.

The temperature and chemical potential dependence of fluctuations and correlations of conserved charges can be useful tools serving as indicators for the transition behavior. The fluctuations and correlations of the charges are given respectively by:
\begin{align}
\chi^{i}_{n}&\equiv \frac{1}{n!}\frac{\partial^{n}\Omega/T^4}{\partial \left(\frac{\mu_i}{T}\right)^n}\nonumber\\
\chi^{i,\,j}_{n,\,m}&\equiv \frac{1}{n!}\frac{1}{m!}\frac{\partial^{n+m}\Omega/T^4}{\left(\partial\frac{\mu_j}{T}\right)^n\left(\partial\frac{\mu_i}{T}\right)^m}
\end{align}
Here we will consider the results pertaining to baryonic number $N_B$, electric charge number $N_Q$, and  strangeness number $N_S$. The corresponding chemical potentials are related to $\mu_i$ $i=u,\,d,\,s$ through $\mu_Q=\mu_u-\mu_d$, $\mu_B=2\mu_d+\mu_u$ and $\mu_S=\mu_d-\mu_s$ \footnote{As can be readily deduced from the relations of the corresponding numbers: $N_B=\frac{1}{3}\left(N_u+N_d+N_s\right)$, $N_Q=+\frac{2}{3}N_u-\frac{1}{3}\left(N_d+N_s\right)$ and $N_S=-N_s$ (by convention strangeness number is the negative of the number of strange quarks).}.

The traced energy-momentum tensor $\Theta^\mu_{~\mu}$ and the speed of sound $C_s$ are also thermodynamical quantities of interest. Both of these have been evaluated in lQCD thus we can use them as benchmarks to evaluate the adequacy of our models. They can be obtained respectively as:
\begin{align}
\Theta^\mu_{~\mu}&=\epsilon-3 P\nonumber\\
C^2_s&\equiv \frac{\partial P}{\partial \epsilon}=\frac{s}{C_V}=\frac{-\frac{\partial\Omega}{\partial T}}{T\frac{\partial^2\Omega}{\partial T^2}},
\end{align}
with $P$ denoting the pressure, $\epsilon=T s- P$ the energy density, $s=\partial P/\partial T$ the entropy density and $C_V=(\partial\epsilon/\partial T)_V$ the specific heat at constant volume.
\subsection{Polyakov loop extension}
\label{pol}
Although no gluonic degrees of freedom are present in the NJL model its extension to the so called Polyakov--Nambu--Jona-Lasinio Model %\cite{Fukushima:2003fw,Megias:2003ui,Megias:2004hj,Roessner:2006xn},%\cite{Ghosh:2007wy,Fu:2007xc,Costa:2008dp,Fukushima:2008wg,Moreira:2010bx,Bhattacharyya:2010wp,Stiele:2016cfs,Bhattacharyya:2016jsn}
 is an attempt to mimic part of its dynamics by considering a static homogeneous background gluonic field in the temporal gauge $A_4=i A^0$ which is diagonal in color space $A_4=A_3\lambda_3+A_8\lambda_8$ and couples with strength $g$ to the quark fields through the covariant derivative ${\cal D}^\mu= \partial^\mu + i A^\mu$, $A^\mu=\delta_0^\mu g A_a^0 \lambda^a/2$, where $\lambda_a$ are the Gell-Mann matrices in color space. 

The Polyakov loop $L$, winding around the imaginary time with periodic boundary conditions, and its trace in color space, $\phi$ (and charge conjugate $\bar\phi$)  are given as 
\begin{align}
L&=\mathcal{P}e^{\int^\beta_0\left(\imath A_4\right)\mathrm{d} x_4}\nonumber\\
\phi&=\frac{1}{N_c}\mathrm{Tr\left[L\right]},\quad\bar{\phi}=\frac{1}{N_c}\mathrm{Tr\left[L^\dag\right]},
\end{align}
where  ${\cal P}$ stands for path-ordering and $\beta=1/T$. In the quenched limit $\phi$ is an order parameter  for the transition between the confined phase where the center of $SU(N_c)$ symmetry ($\mathbb{Z}_{N_c}$) is verified (vanishing traced Polyakov loop) and the deconfined phase where this symmetry is spontaneously broken \cite{McLerran:1981pb}.

 An additional term, the Polyakov potential, must be added to drive this temperature induced spontaneous breaking.  Its form can be determined by fitting lattice QCD observables. We take two choices, ${\cal U}_I$ \cite{Roessner:2006xn} and ${\cal U}_{II}$ \cite{Ghosh:2007wy,Bhattacharyya:2016jsn}, with parameters shown in Table \ref{pnjltab},

\begin{itemize}
\item Logarithmic form 
%\footnote{\tiny S. Rößner, C. Ratti, and W. Weise, Phys. Rev. D 75, 034007 }\\
%($a_0=3.51,~a_1=-2.47,~a_2=15.2,~b_3=-1.75$)
%:
\begin{align}
\label{UPI}
&\frac{\mathcal{U}_{I}\left[\phi,\bar{\phi},T\right]}{T^4}\nonumber\\
=&-\frac{1}{2}a\left(T\right)\bar{\phi}\phi \nonumber\\
&
+b\left(T\right)\mathrm{ln}\left[1-6\bar{\phi}\phi+4\left(\bar{\phi}^3+\phi^3\right)-3\left(\bar{\phi}\phi\right)^2\right]\nonumber\\
a\left(T\right)&=
%\left(
a_0+a_1\frac{T_0}{T}+a_2\left(\frac{T_0}{T}\right)^2\nonumber\\
%\right)
b\left(T\right)&=b_3 \left(\frac{T_0}{T}\right)^3
\end{align}
\item Exponential K-Log form 
%\footnote{\tiny A. Bhattacharyya, S. K. Ghosh, S. Maity, S. Raha, Rajarshi Ray, K. Saha, and S. Upadhaya, Phys. Rev. D 95, 054005 }\\
%($a_0=6.75,~a_1=-9.8,~a_2=0.26,~b_3=0.805,~b_4=7.555,~K=0.1 $):
\begin{align}
\label{UPII}
&\frac{\mathcal{U}^\prime_{II}\left[\phi,\bar{\phi},T\right]}{T^4}\nonumber\\
=&-\frac{1}{2}a\left(T\right)\bar{\phi}\phi
-\frac{b_3}{6}\left(\bar{\phi}^3+\phi^3\right)
+\frac{b_4}{4}\left(\bar{\phi}\phi\right)^2 \nonumber\\
&+
K\mathrm{ln}\left[
\frac{27}{24 \pi^2}\left(
1-6\bar{\phi}\phi+4\left(\bar{\phi}^3+\phi^3\right)-3\left(\bar{\phi}\phi\right)^2
\right)
\right]\nonumber\\
a\left(T\right)&=
a_0+a_1\left(\frac{T_0}{T}\right) e^{-a_2\frac{T_0}{T}}
\end{align}
\end{itemize}
where the term proportional to $K$ is the Van der Monde determinant, and the exponential term going with $a_2$ is a modification introduced in   \cite{Bhattacharyya:2016jsn}. We considered a slight modification of this potential as we used $\mathcal{U}_{II}\left[\phi,\bar{\phi},T\right]=\mathcal{U}^\prime_{II}\left[\phi,\bar{\phi},T\right]-\mathcal{U}^\prime_{II}\left[0,0,T\right]$ which enables the reproduction of the expected vanishing value for $\Omega/T^4$ as we approach the vacuum ($\left\{T,\mu\right\}=\left\{0,0\right\}$).

From a practical point of view the extension from NJL to PNJL amounts to the introduction of two new classical fields in the model, $\phi$ and $\bar{\phi}$, the introduction of the Polyakov potential and a modification of the occupation numbers (see for instance \cite{Moreira:2010bx} for details on the implementation of the model) \footnote{The easiest way to introduce this modification is to note that the phase of the Polyakov loop appears in the quark action in the form of an imaginary chemical potential \cite{Weiss:1980rj,Weiss:1981ev,Fukushima:2003fw}.}.  At vanishing baryonic chemical potential, the case considered in this work, $\phi=\bar \phi$.

\subsection{Parameter fitting}
\label{fits}
The parameters of the model are the quark current masses ($m_u$, $m_d$ and $m_s$), the cutoff ($\Lambda$) and the couplings ($G$, $\kappa$, $\kappa_2$, $g_1$, $g_2$, $g_3$, $g_4$, $g_5$, $g_6$, $g_7$ and $g_8$). In the formulation of the model without non-canonical explicit chiral symmetry breaking, NJLH8q, these $15$ parameters are reduced to $8$ ($\kappa_2$ and $g_i$ with $i=3,\ldots,8$ are set to zero). If we neglect the isospin symmetry breaking ($m_l\equiv m_u=m_d$) these are further reduced to $7$. 

As was shown in \cite{Osipov:2006ns} the NJLH8q model can be fitted for several fixed values of the OZI violating eight-quark interaction coupling $g_1$ (the four-quark interaction strength, $G$, is smaller for increasing $g_1$ but the remaining parameters are unchanged) while keeping the mesonic spectra unchanged apart from a decrease in the $\sigma$ meson mass for increasing value of $g_1$ (the scalar mixing angle also changes). Two sets, denoted as NJLH8qA and NJLH8qB, are shown in Table \ref{ParameterSets} with the latter corresponding to the highest value of $g_1$ (and conversely the lowest $G$). In these isospin symmetric sets ($m_l\equiv m_u=m_d$) we fit the model parameters by imposing a value of $g_1$ and fitting the remaining 6 parameters using the pion and kaon weak decay couplings ($f_\pi$ and $f_K$) and the %pion, kaon, eta-prime and $a_0$ 
meson masses ($M_\pi$, $M_K$, $M_{\eta^\prime}$ and $M_{a_0}$).

This freedom allowed us to isolate and study the impact of the eight-quark interaction term in the model phase diagram in the chemical potential-temperature, $\left\{\mu,T\right\}$, plane  \cite{Hiller:2008nu}. One of the main highlights of this study was the realization that the CEP is shifted to lower chemical potential and higher temperature with increasing $g_1$. This in turn leads to a substantial reduction in  the related crossover temperature at $\mu=0$  compared to the case with weak $g_1$ coupling, as reported before in \cite{Osipov:2006ev,Osipov:2007mk}, with the lower values of T  complying with lQCD results \cite{Aoki:2006br}.

The extension to include the  non-canonical explicit chiral-symmetry breaking interactions introduces 7 new parameters. For the parameter set NJLH8qmA from Table \ref{ParameterSets}  we chose to impose the value of the current masses ($m_l$ and $m_s$). The remaining 12 parameters can be fitted by fixing $f_\pi$, $f_K$, the pseudoscalar and scalar mixing angles ($\theta_{ps}$ and $\theta_s$) and the 8 meson masses ($M_\pi$, $M_K$, $M_\eta$, $M_{\eta^\prime}$, $M_{a_0}$, $M_{K^\ast}$, $M_\sigma$ and $M_{f_0}$), in the isospin limit. Note that the inclusion of the ESB interactions allows to fit the pseudoscalar as well as the scalar spectra to empirical data with a high degree of accuracy, as well as the weak decay constants and the current quark mass values.  This on the other hand reduces the former freedom in the interplay of $G,g_1$ parameters, which is now considerably narrowed down, favoring the strong $g_1$ coupling strength,  (we are considering here a range $470~\mathrm{MeV}\lessapprox M_\sigma\lessapprox 500~\mathrm{MeV}$). One also sees that the increase in $g_1$ comes in this case accompanied not only by a decrease in $G$, but also a decrease in the ESB couplings $g_4,g_7$.

\begin{table*}
\caption{Model parameters obtained using a regularization kernel with two Pauli-Villars subtractions in the integrand (see \cite{Osipov:1985})
 given in the following units: for the current masses $\left[m_i\right]=\text{MeV}$ ($i=l,\,u,\,d,\,s$), for the cutoff $\left[\Lambda\right]=\text{MeV}$, for the couplings $\left[G\right]=\text{GeV}^{-2}$, $\left[\kappa_2\right]=\text{GeV}^{-3}$, $\left[g_5\right]=\left[g_6\right]=\left[g_7\right]=\left[g_8\right]=\text{GeV}^{-4}$,  $\left[\kappa\right]=\text{GeV}^{-5}$ and $\left[g_3\right]=\left[g_4\right]=\text{GeV}^{-6}$, $\left[g_1\right]=\left[g_2\right]=\text{GeV}^{-8}$. Parameters marked with an asterisk ($^\ast$) were kept fixed. 
Several quantities which are either outputs or kept fixed (and used in the fit of the remaining parameters) are presented in the bottom rows: weak decay couplings ($[f_\pi]=[f_K]=\text{MeV}$), meson masses for the low-lying scalars/pseudo-scalars, the dynamical masses of the quarks  (given in $\text{MeV}$) and the corresponding chiral condensates $\langle\bar{q}q\rangle_i$ (given in $\text{MeV}^3$) taking into account the subtraction of the contribution coming from the current mass, see Eq. \ref{qco}. The pseudoscalar and scalar mixing angles ($\theta_{ps}$/$\theta_{s}$) 
 are given in degrees.
Sets NJLH8qA and NJLH8qB include up to eight-quark interactions but no non-canonical explicit chiral symmetry breaking terms whereas set NJLH8qmA does include these terms. 
%
%In the set NJLH8qmIBrA we consider different masses for the up and down quarks thus breaking isospin symmetry.
%
}
\label{ParameterSets}
\begin{footnotesize}
%\begin{tiny}
%parametros sem isobreak
\begin{tabular*}{\textwidth}{@{\extracolsep{\fill}}lrrrrrrrrrrrrrr@{}}\hline
  \multicolumn{1}{c}{Set}
& \multicolumn{1}{c}{$m_l$}
&	\multicolumn{1}{c}{$m_s$}
& \multicolumn{1}{c}{$G$}
& \multicolumn{1}{c}{$\kappa$}
& \multicolumn{1}{c}{$\kappa_2$}
& \multicolumn{1}{c}{$g_1$}
& \multicolumn{1}{c}{$g_2$}
& \multicolumn{1}{c}{$g_3$}
& \multicolumn{1}{c}{$g_4$}
& \multicolumn{1}{c}{$g_5$}
& \multicolumn{1}{c}{$g_6$}  
& \multicolumn{1}{c}{$g_7$}  
& \multicolumn{1}{c}{$g_8$}  
& \multicolumn{1}{c}{$\Lambda$}\\
\hline
NJLH8qA & $5.94$ & $186.12$  & $10.92$ & $-125.07$ & $0^\ast$ & $500^\ast$ & $-47.14$ & $0^\ast$ & $0^\ast$   & $0^\ast$  & $0^\ast$    & $0^\ast$  & $0^\ast$  & $0.851$\\
NJLH8qB & $5.94$ & $186.12$  &  $8.14$ & $-125.07$ & $0^\ast$ & $3000^\ast$ & $-47.14$ & $0^\ast$ & $0^\ast$   & $0^\ast$  & $0^\ast$    & $0^\ast$  & $0^\ast$  & $0.851$\\
NJLH8qmA & $4^\ast$ & $100^\ast$ & $10.08$ & $-114.25$ & $6.00$  & $3641.45$ & $49.42$  & $-4313.92$ & $1589.24$ & $190.53$ & $-1171.23$ & $163.28$ & $-60.79$ & $0.838$\\
NJLH8qmB & $4^\ast$ & $100^\ast$ &  $9.35$ & $-114.25$ & $6.00$  & $3976.88$ & $49.42$  & $-4313.92$ & $1320.29$ & $190.53$ & $-1171.23$ & $116.31$ & $-60.79$ & $0.838^\ast$
\end{tabular*}
%observaveis sem isobreak I
\begin{tabular*}{\textwidth}{@{\extracolsep{\fill}}lrrrrrrrr@{}}\hline
  \multicolumn{1}{c}{Set}
& \multicolumn{1}{c}{$f_\pi$}
& \multicolumn{1}{c}{$f_K$}  
& \multicolumn{1}{c}{$\theta_{ps}$}
& \multicolumn{1}{c}{$\theta_{s}$}
& \multicolumn{1}{c}{$M_l$}
& \multicolumn{1}{c}{$M_s$}
& \multicolumn{1}{c}{$\langle\bar{q} q\rangle_l$}
& \multicolumn{1}{c}{$\langle\bar{q} q\rangle_s$}
\\\hline
NJLH8qA  &	$92^\ast$ & $117^\ast$ & $-13.98$   & $23.29$      & $359.19$ & $554.40$ & $-(233.88)^3$ & $-(182.76)^3$ \\
NJLH8qB  &	$92^\ast$ & $117^\ast$ & $-13.98$   & $19.71$      & $359.19$ & $554.40$ & $-(233.88)^3$ & $-(182.76)^3$ \\
NJLH8qmA &	$92^\ast$ & $113^\ast$ & $-12^\ast$ & $27.50^\ast$ & $360.10$ & $524.49$ & $-(231.43)^3$ & $-(208.51)^3$	\\
NJLH8qmB &	$92^\ast$ & $113^\ast$ & $-12^\ast$ & $25.80$      & $360.10$ & $524.49$ & $-(231.43)^3$ & $-(208.51)^3$
\end{tabular*}
%observaveis sem isobreak I
\begin{tabular*}{\textwidth}{@{\extracolsep{\fill}}lrrrrrrrr@{}}\hline
  \multicolumn{1}{c}{Set}
& \multicolumn{1}{c}{$M_\pi$}
& \multicolumn{1}{c}{$M_K$}
& \multicolumn{1}{c}{$M_\eta$}
& \multicolumn{1}{c}{$M_{\eta^\prime}$}
& \multicolumn{1}{c}{$M_{a_0}$}
& \multicolumn{1}{c}{$M_{K^\star}$}
& \multicolumn{1}{c}{$M_\sigma$}
& \multicolumn{1}{c}{$M_{f_0}$}\\\hline
NJLH8qA  & $138^\ast$ & $494^\ast$ & $477.50$ & $958^\ast$ & $980^\ast$ & $1200.93$ & $691.17$ & $1368.04$\\
NJLH8qB  & $138^\ast$ & $494^\ast$ & $477.50$ & $958^\ast$ & $980^\ast$ & $1200.93$ & $520.83$ & $1352.94$\\
NJLH8qmA & $138^\ast$ & $494^\ast$ & $547^\ast$ & $958^\ast$ & $980^\ast$ & $890^\ast$ & $500^\ast$ & $980^\ast$\\
NJLH8qmB & $138^\ast$ & $494^\ast$ & $547^\ast$ & $958^\ast$ & $980^\ast$ & $890^\ast$ & $480^\ast$ & $980^\ast$\\
\hline
\end{tabular*}
\end{footnotesize} 
\end{table*}

\begin{table*}
\caption{The parameters for the Polyakov potentials ${\cal U}_I$ and ${\cal U}_{II}$  given in Eqs. \ref{UPI} and \ref{UPII}. 
The parameter $T_0$ sets the temperature scale at which deconfinement arises. The main effect of a modification of this parameter is that of shifting the transitional temperatures towards higher/lower temperatures with larger/smaller $T_0$. While $T_0=270~\mathrm{MeV}$ is given in \cite{Roessner:2006xn}
as the value stemming from lQCD calculations in pure gauge, it is expected that this temperature should be adjusted to reflect the inclusion of dynamical quarks and the number of flavors considered \cite{Schaefer:2007pw, Schaefer:2009ui}. Here we chose values for $T_0$ that gave a closer agreement to the lQCD data.
} 
\label{pnjltab}
\begin{footnotesize}
%\begin{tiny}
%parametros sem isobreak
\begin{tabular*}{\textwidth}{@{\extracolsep{\fill}}lrrrrrrr@{}}\hline
\multicolumn{1}{c}{${\cal U}_1$\cite{Roessner:2006xn}}%,\cite{Stiele:2016cfs}
& \multicolumn{1}{c}{$T_0$ [MeV]}
& \multicolumn{1}{c}{$a_0$}
&\multicolumn{1}{c}{$a_1$}
& \multicolumn{1}{c}{$a_2$}
& \multicolumn{1}{c}{$b_3$}
& \multicolumn{1}{c}{}
& \multicolumn{1}{c}{}\\
%
%\hline
&200 &  $0.351$ & $-2.47$  & $ 15.2$ & $ -1.75$ &  &      \\
\hline
${\cal U}_{II}$ \cite{Ghosh:2007wy,Bhattacharyya:2016jsn} & $T_0$ [MeV] & $a_0$ &  $a_1$ & $a_2$ & $b_3$  & $b_4$ & $K$   \\
& $175$ & $6.75$ &  $-9.8$ & $0.26$ & $0.805$  & $7.555$ & $0.1$    \\
\hline
\end{tabular*}
\end{footnotesize} 
\end{table*}
%($a_0=3.51,~a_1=-2.47,~a_2=15.2,~b_3=-1.75$)
%($a_0=6.75,~a_1=-9.8,~a_2=0.26,~b_3=0.805,~b_4=7.555,~K=0.1 $)

\section{Results and discussion}
\label{Results}
%\subsection{\texorpdfstring{$T-\mu$}{T-mu} phase diagram}
%\label{PhaseDiagram}
\subsection{Speed of sound and energy-momentum trace anomaly}
\label{CS2TraAnom}

\begin{figure*}
\center
\subfigure[]{\label{Cs2ELAPV}\includegraphics[width=0.32\textwidth]{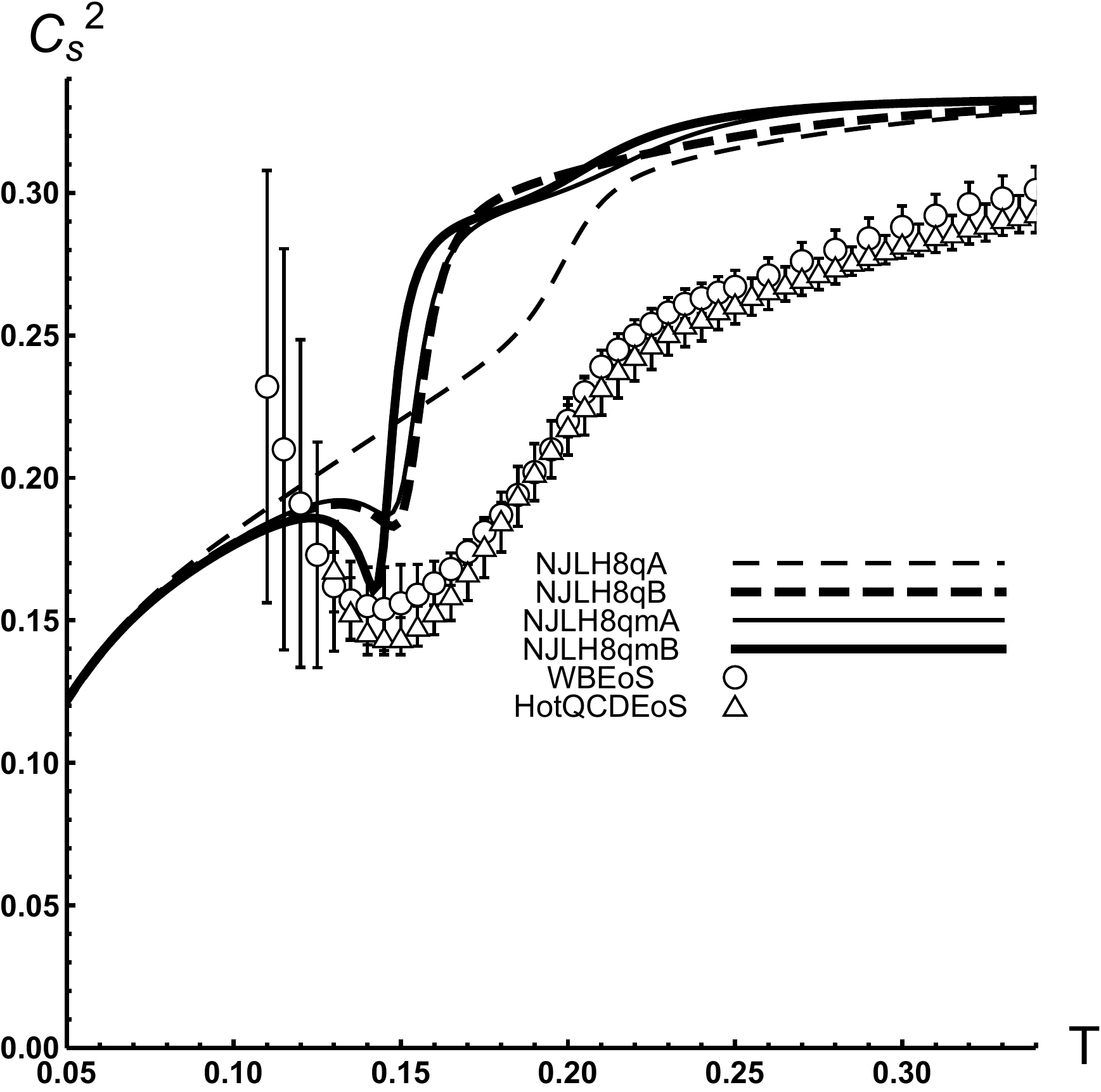}}
\subfigure[]{\label{TmumuELAPV}\includegraphics[width=0.32\textwidth]{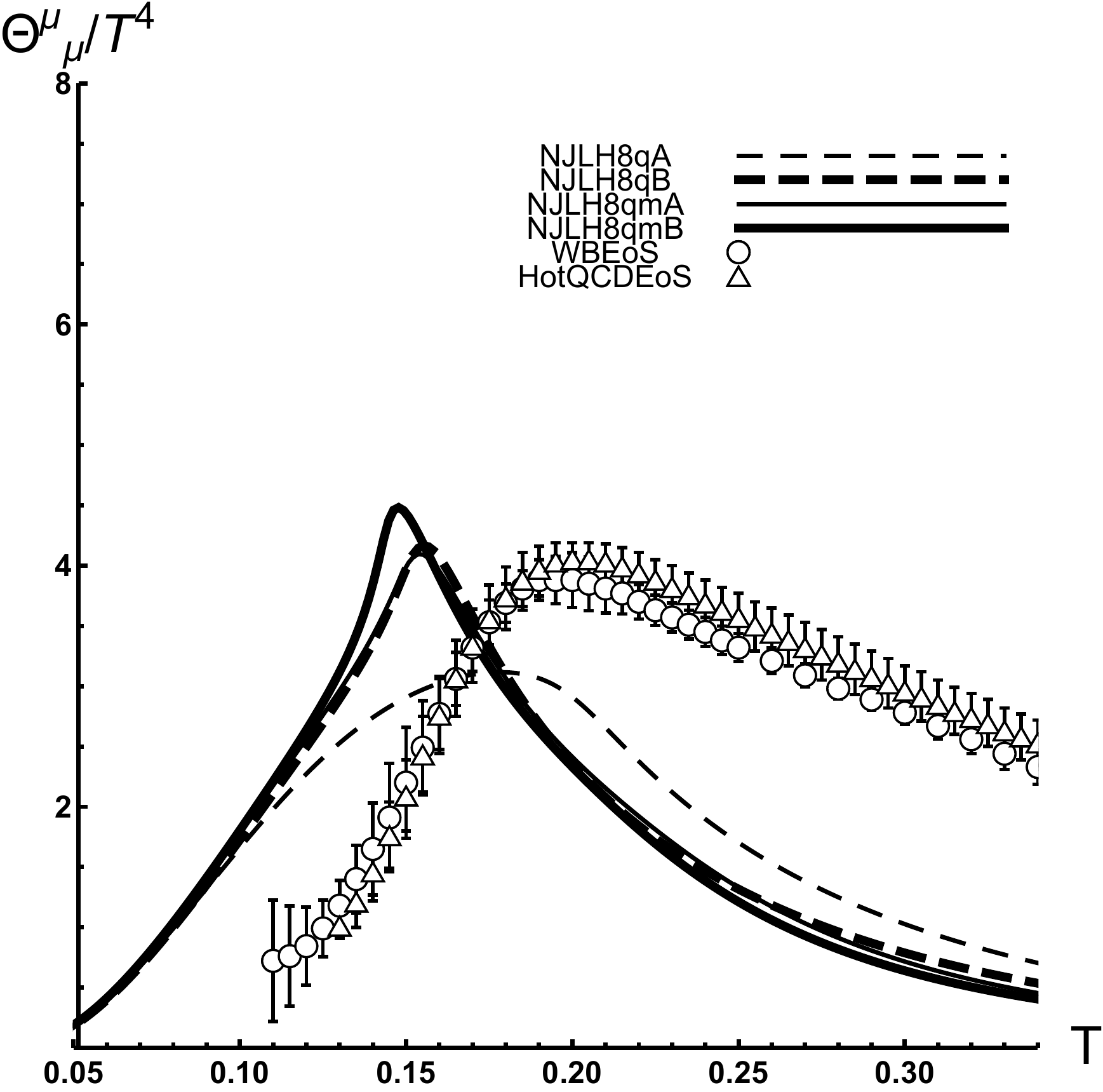}}
\caption{The squared speed of sound (\ref{Cs2ELAPV}) and the energy-momentum trace anomaly (\ref{TmumuELAPV} ) as  functions of the temperature ($[T]=\mathrm{GeV}$) at vanishing chemical potential, obtained using the parameter sets from  Table \ref{ParameterSets}: dashed lines correspond to the NJLH8q sets whereas solid lines correspond to the NJLH8qm sets (respectively the ones without and with non-canonical explicit chiral symmetry breaking).
 The markers, labeled as WBEoS and HotQCDEoS, correspond to continuum extrapolated lQCD results taken respectively from \cite{Borsanyi:2013bia} and \cite{Bazavov:2014pvz}. 
}
\label{Cs2TrAnomELAPV}
\end{figure*}

In Fig. \ref{Cs2ELAPV} we see a comparison of the temperature dependence of the squared speed of sound at vanishing chemical potential as obtained using lQCD and several parametrizations of our model.  Although a complete quantitative agreement seems impossible with the considered parametrizations an approximation of the general qualitative behavior can be obtained: the squared speed of sound increases with temperature starting at zero until a critical point upon which it dips into a local minimum (for 3 of the sets) and then rises again going asymptotically to the Stefan-Boltzmann limit. For temperatures above this local minimum the obtained speed of sound overshoots the lQCD result whereas for lower temperatures we obtain lower values. Note that as one expects the speed of sound to go to zero in the limit of vanishing temperature and chemical potential the lQCD results should be followed by a dip towards zero for lower temperatures.

The model velocity of sound displays a soft point close to the one of lQCD, at around $T\sim 150$ MeV. For this feature to be present in the model, the parameter $g_1$ of OZI-violating 8-quark interactions is required to have a certain strength $g_1\sim 3000-4000$ GeV$^{-8}$, see set NJLH8qB without ESB interactions, and  sets NJLH8qmA, NJLH8qmB with ESB in Table \ref{ParameterSets}; at the weak coupling $g_1=500$ GeV$^{-8}$ of set NJLH8qA  the relative minimum is absent and the velocity of sound shows a monotonous decrease. It had already been noticed some time ago that the strength of parameter $g_1$ has impact on the number of degrees of freedom \cite{Hiller:2008nu}; a  rather strong (more than 50 $\%$) suppression of the artificial quark excitations (due to lack of confinement of the model)  at $T /T_c > 0$  was observed, in comparison to a PNJL model calculation \cite{Fukushima:2008wg}. Thus it is understandable that this property manifests itself in the occurrence of the relative minimum in the velocity of sound, as the model emulates partially the missing degrees of freedom attributed to the onset of deconfinement. The inclusion of the ESB breaking parameters does not change this important property, in spite of the fact that the accurate fit of the low lying spectra and related properties strongly constrains the parameters of the model in the vacuum. On the contrary: as mentioned before, the ESB interactions together with the requirement of having good fits of the spectra rule out the smaller values of $g_1$ strengths. The fact that the former freedom in the model parameter $g_1$  is narrowed down, and specifically to the values that describe the  soft region in the speed of sound, can be seen as a major result regarding the phenomenological importance of including ESB interactions in the model. 
 
Apart from the relative minimum in the speed of sound, notice that the sets with ESB display a slight dip after the steep rise and before flattening out. This is a remnant of the two CEPs encountered in the model as described previously, and still visible at $\mu_B=0$. This behavior may be guessed to be very subtly present in the lQCD points as well, although it would require further investigation to clarify this point.

Moving to Fig. \ref{TmumuELAPV} , one sees that the height of the peak of the trace of the energy momentum tensor is improved, as well as the slope before the transition, comparing with lQCD data, by the inclusion of the ESB terms (or the selection of the strong $g_1$ coupling without ESB). To understand that this is natural to expect within our model, we recall that the trace of the energy momentum tensor and the number of degrees of freedom $\nu(T)=(90/\pi^2)\frac{P(T)-P(0)}{T^4}$ are closely related
\begin{equation}
\label{MT}
\frac{\Theta^{\mu\mu}}{T^4}=\frac{\pi^2 }{90}(T\frac{\partial}{\partial T} \nu(T)).
\end{equation}
As mentioned above, the number of degrees of freedom for weak and strong $g_1$ coupling was obtained in \cite{Hiller:2008nu,Moreira:2010bx}, as function of $T/T_c$, where $T_c$ is the crossover temperature. Converting by this factor one obtains the slope behavior displayed in Fig.  \ref{TmumuELAPV}.

The slope after the peak is steeper in the model than in lattice calculations, but this is also expected, since the model approaches the Stefan-Boltzmann limit faster. Furthermore since in the non-Polyakov loop extended case there are no gluonic degrees of freedom present, this limit corresponds to a lower value ($\lim_{T\rightarrow\infty}-\Omega_{NJL}/T^4=31.5 \left(\pi^2/90\right)$ whereas $\lim_{T\rightarrow\infty}-\Omega_{PNJL}/T^4=47.5 \left(\pi^2/90\right)$, see for instance \cite{Hiller:2008nu,Moreira:2010bx}).

\begin{figure*}
\center
\subfigure[]{\label{PELAPV}\includegraphics[width=0.32\textwidth]{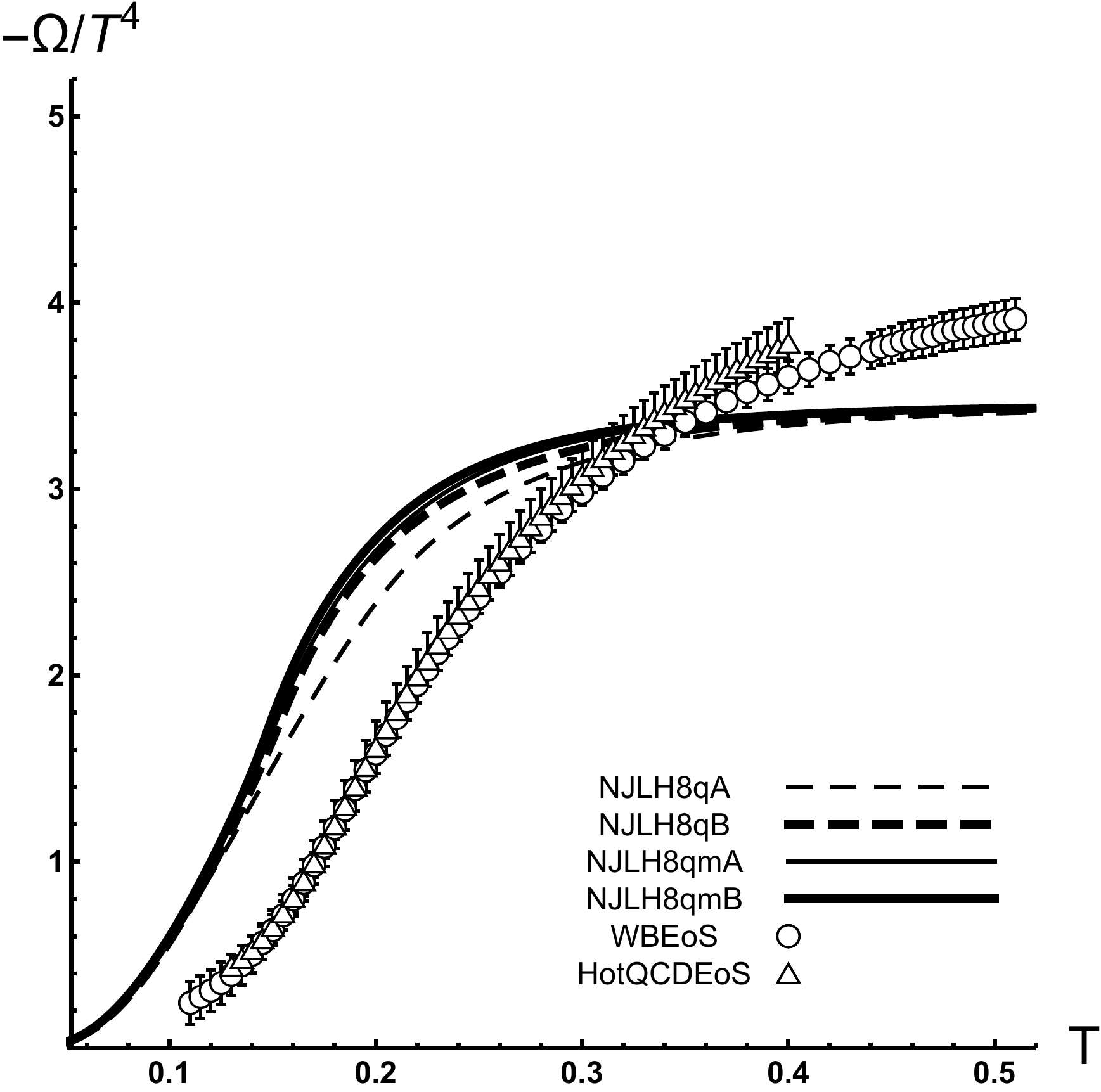}}
\subfigure[]{\label{epsELAPV}\includegraphics[width=0.32\textwidth]{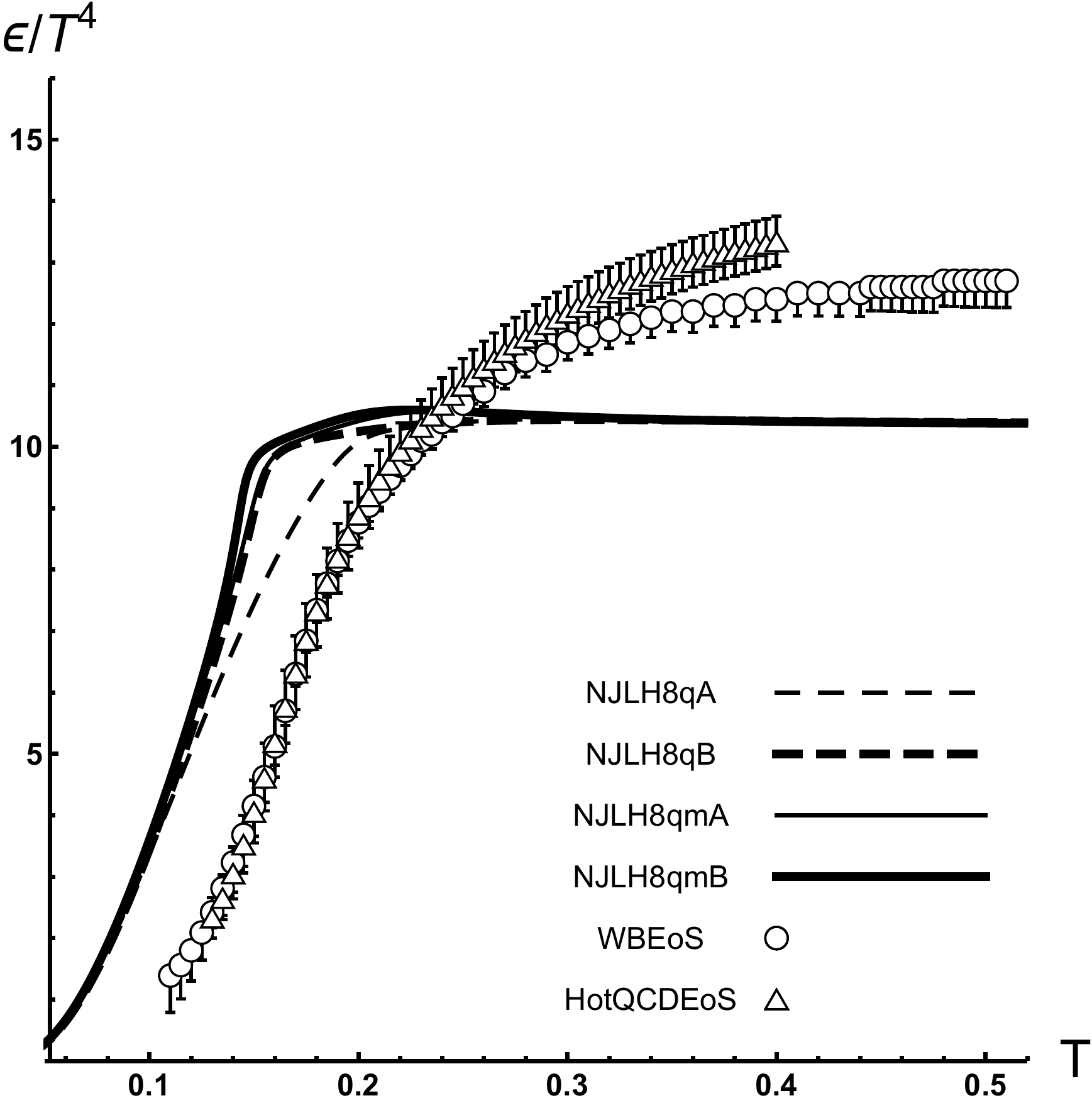}}\\
\subfigure[]{\label{SELAPV}\includegraphics[width=0.32\textwidth]{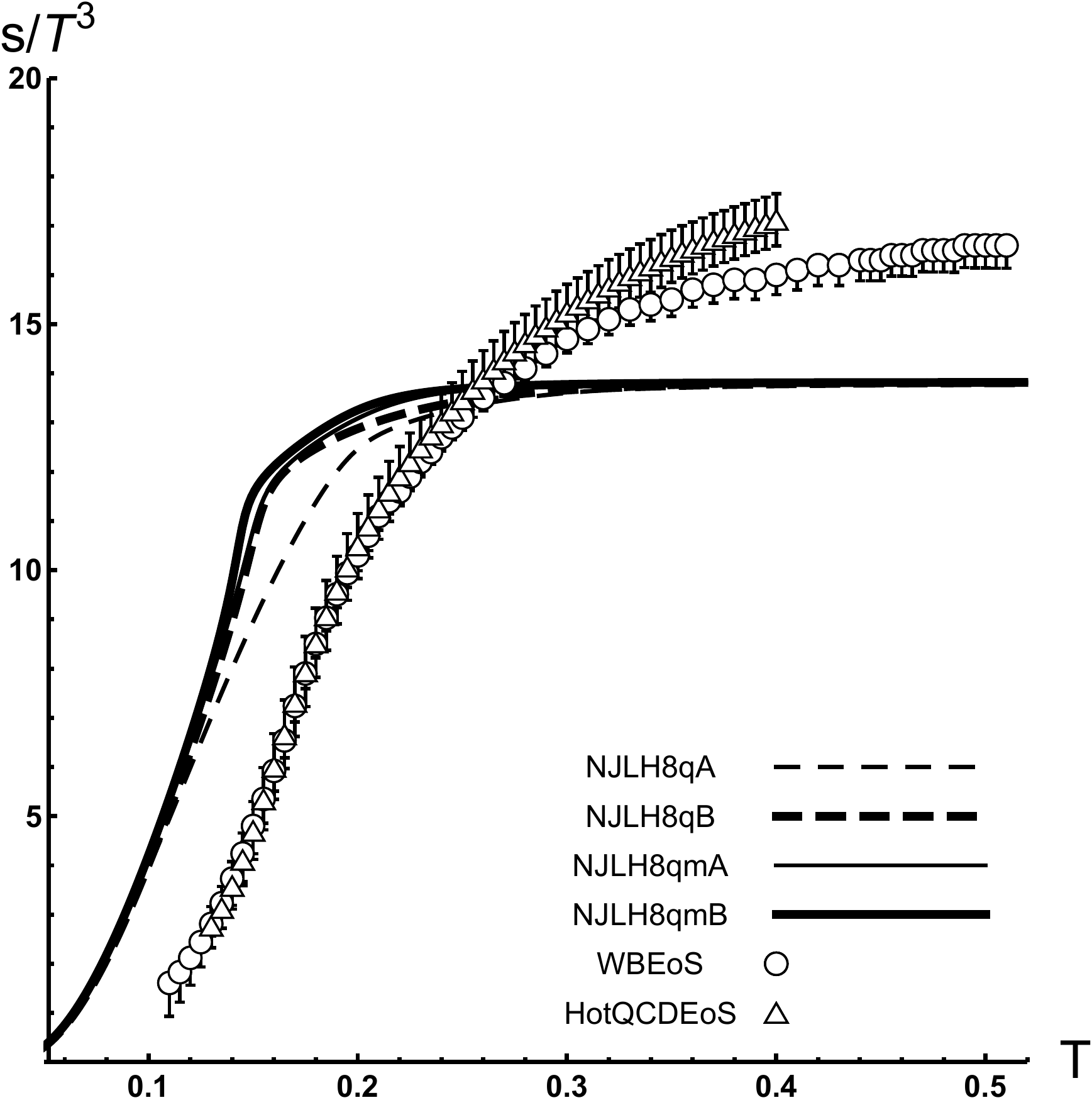}}
\subfigure[]{\label{CVELAPV}\includegraphics[width=0.32\textwidth]{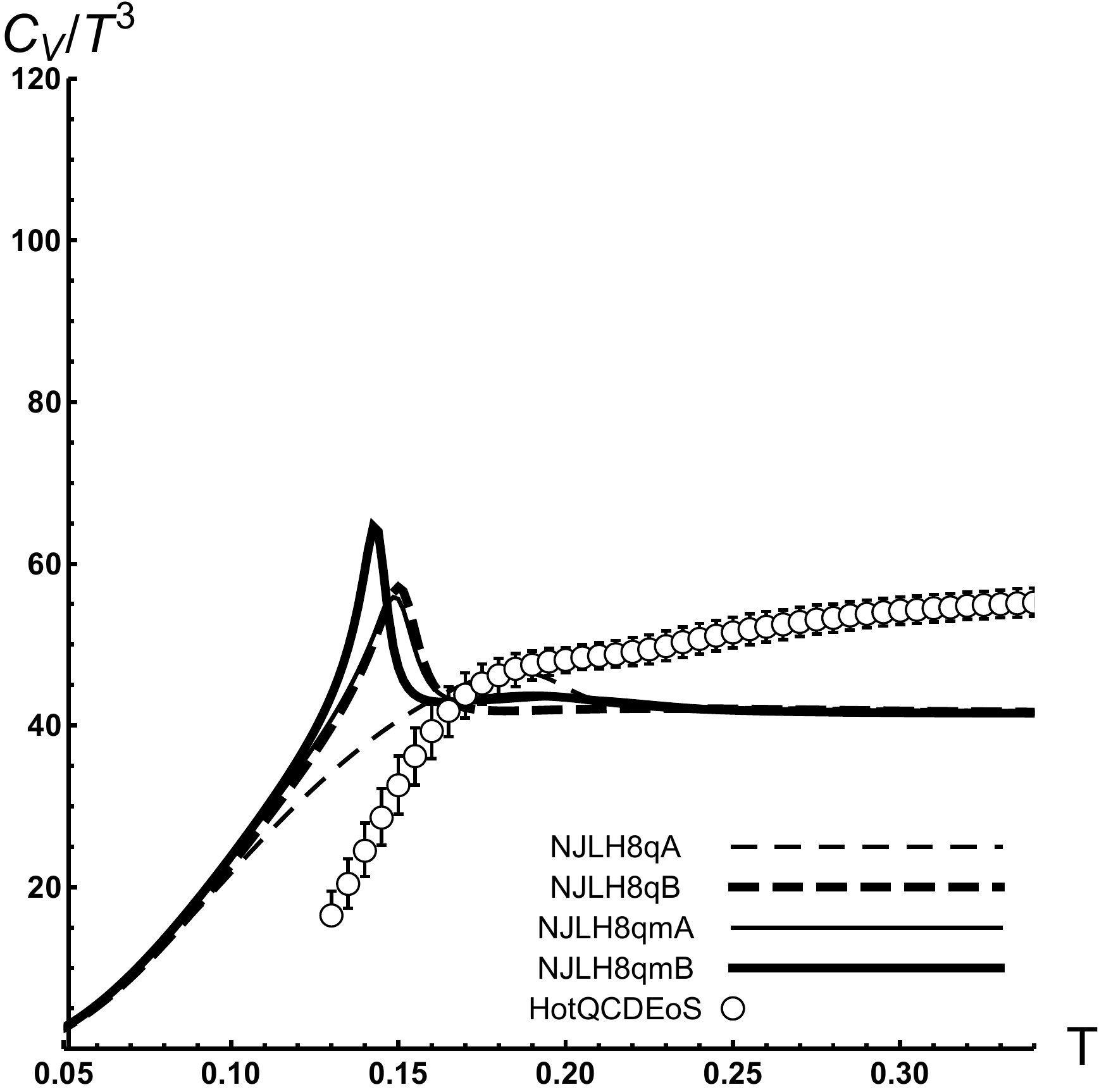}}
\caption{
Dimensionless quantities obtained by dividing the pressure \ref{PELAPV}, energy density \ref{epsELAPV}, entropy \ref{SELAPV} and specific heat \ref{CVELAPV} by adequante powers of the temperature, as functions of temperature for vanishing chemical potential ($[T]=\mathrm{GeV}$) using the parameter sets NJLH8q and NJLH8qm. The markers, labeled as WBEoS and HotQCDEoS, correspond to continuum extrapolated lQCD results taken respectively from \cite{Borsanyi:2013bia} and \cite{Bazavov:2014pvz}.
}
\label{PepsSCVELAPV}
\end{figure*}

The energy density, $\epsilon$, and pressure, $P$, as well as their derivatives with respect to the temperature, $T$, the specific heat, $C_V$ and entropy density, $s$, are depicted in Figs \ref{PepsSCVELAPV}. Despite the reasonable agreement, apart from a shift towards lower temperatures, when we look at the energy-momentum trace anomaly, ${\Theta^\mu}_\mu$, (see Fig. \ref{TmumuELAPV}), in the individual thermodynamical quantities involved, $P$ and $\epsilon$, as well as $s$ (See Figs. \ref{PELAPV}, \ref{epsELAPV} and \ref{SELAPV}), the effect of the missing degrees of freedom is clearly present in their asymptotic behavior. In the cases with stronger eight-quark interactions (NJLH8qB, NJLH8qmA/B) the specific heat (see Fig. \ref{CVELAPV}) deviates from the lQCD with a marked peak around the transition region reflecting the faster transitional behavior. The slope of the curve for temperatures lower than the transition is however better reproduced in the cases with stronger $g_1$.

\subsection{Fluctuations and correlations of conserved charges}
\label{FluctCorrel}

\begin{figure*}
\center
\subfigure[]{\label{Chi2BELAPV}\includegraphics[width=0.32\textwidth]{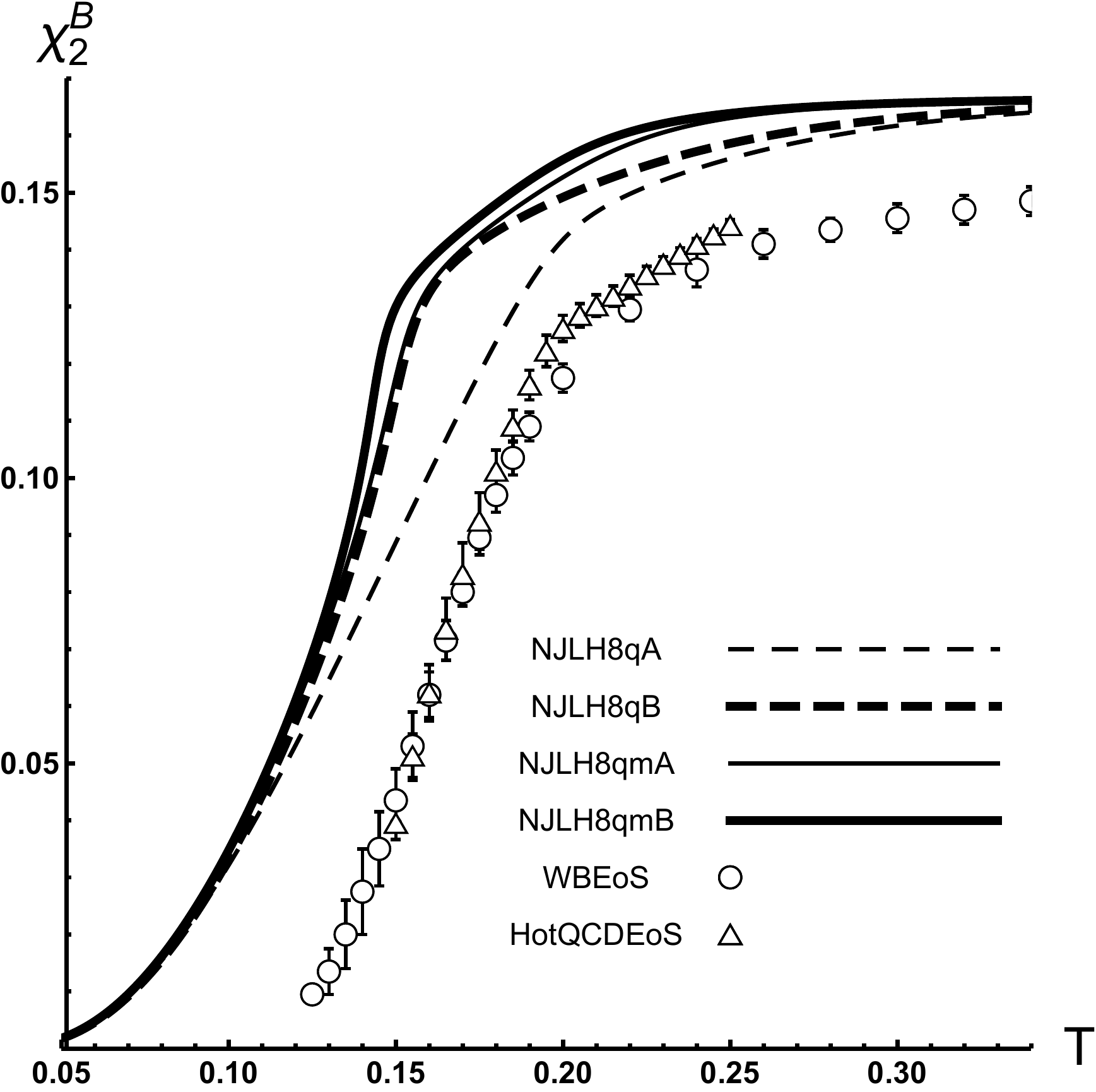}}
\subfigure[]{\label{Chi2QELAPV}\includegraphics[width=0.32\textwidth]{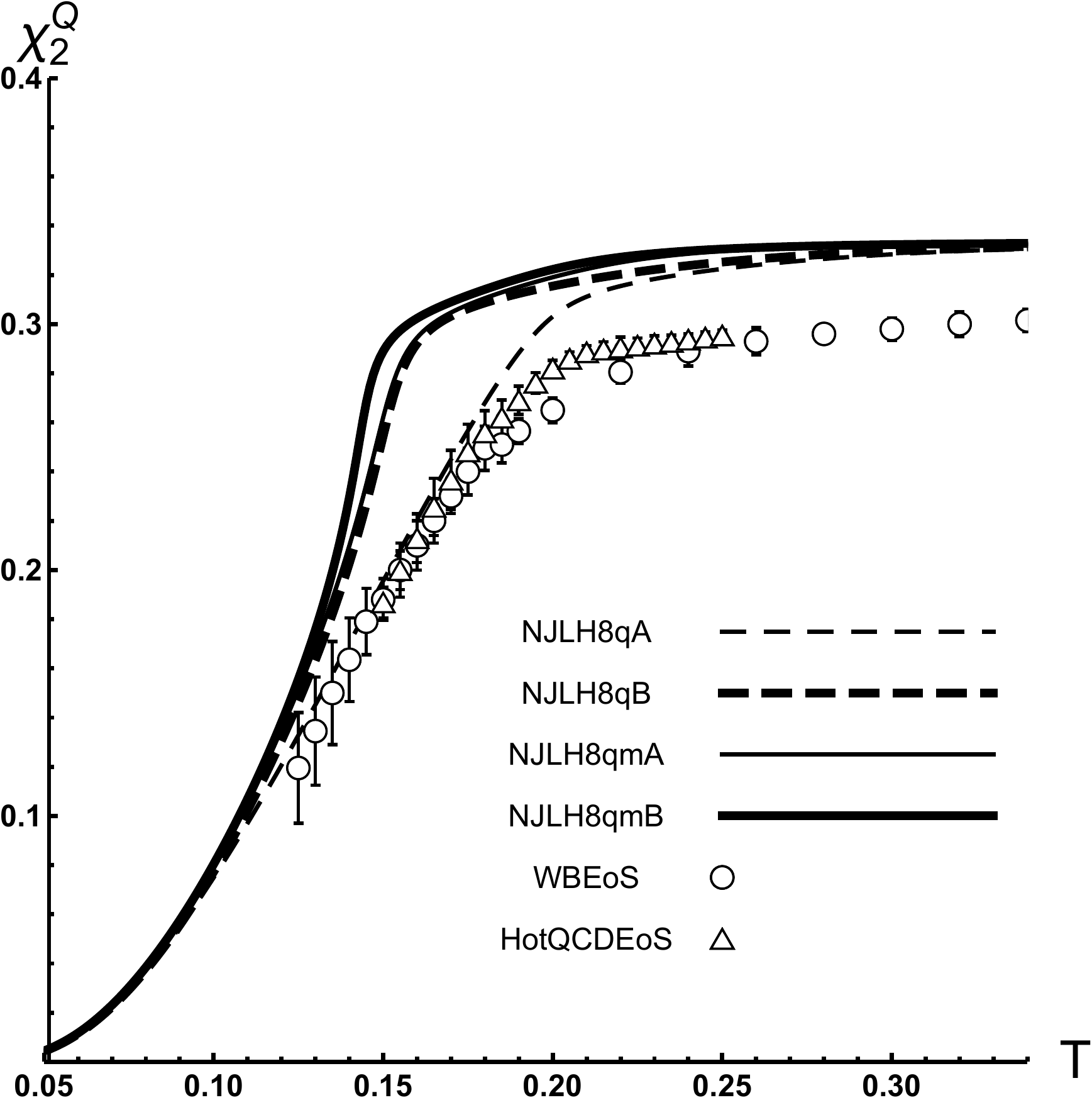}}
\subfigure[]{\label{Chi2SELAPV}\includegraphics[width=0.32\textwidth]{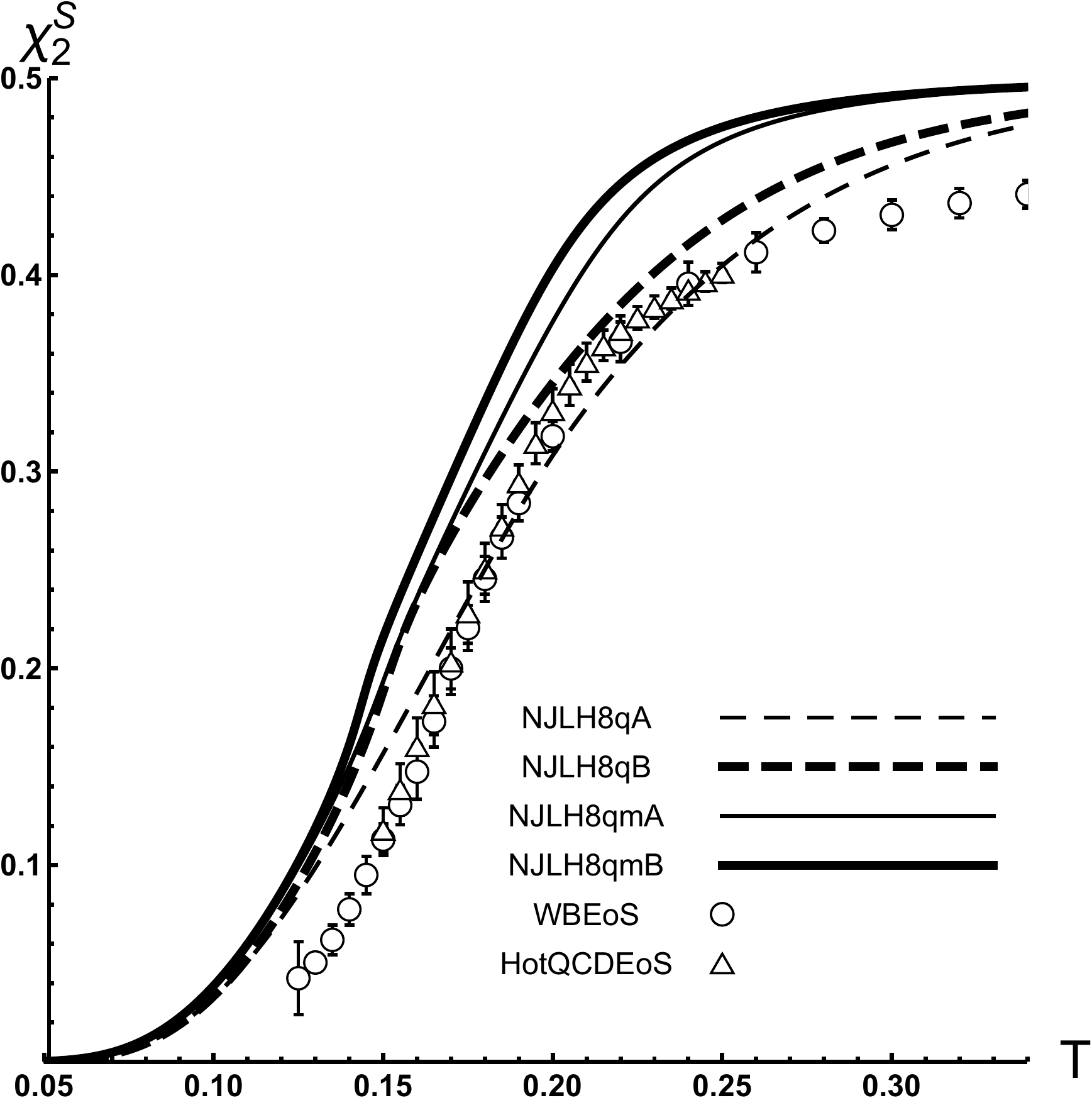}}
\caption{
Fluctuations of conserved charges as functions of temperature ($[T]=\mathrm{GeV}$) at vanishing chemical potential compared to lQCD results: in \ref{Chi2BELAPV} fluctuation of baryonic number ($\chi^B_2$), in \ref{Chi2QELAPV} electric charge ($\chi^Q_2$) and in \ref{Chi2SELAPV} strangeness ($\chi^S_2$).
The markers, labeled as WBEoS and HotQCDEoS, correspond to continuum extrapolated lQCD results taken respectively from \cite{Borsanyi:2011sw} and \cite{Bazavov:2012jq}. 
}
\label{FluctuationsELAPV}
\end{figure*}
\begin{figure*}
\center
\subfigure[]{\label{Chi2BQELAPV}\includegraphics[width=0.32\textwidth]{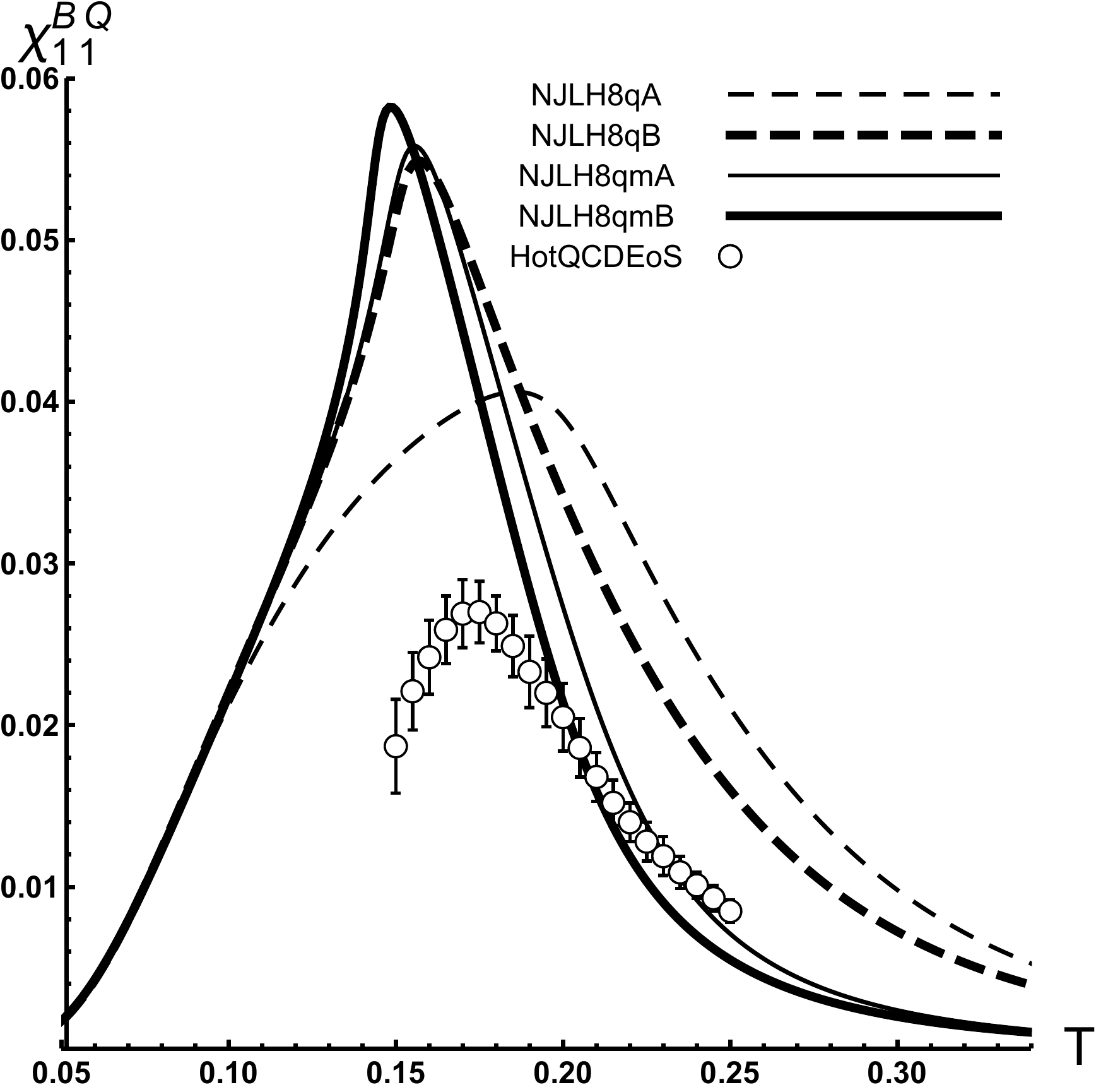}}
\subfigure[]{\label{Chi2BSELAPV}\includegraphics[width=0.32\textwidth]{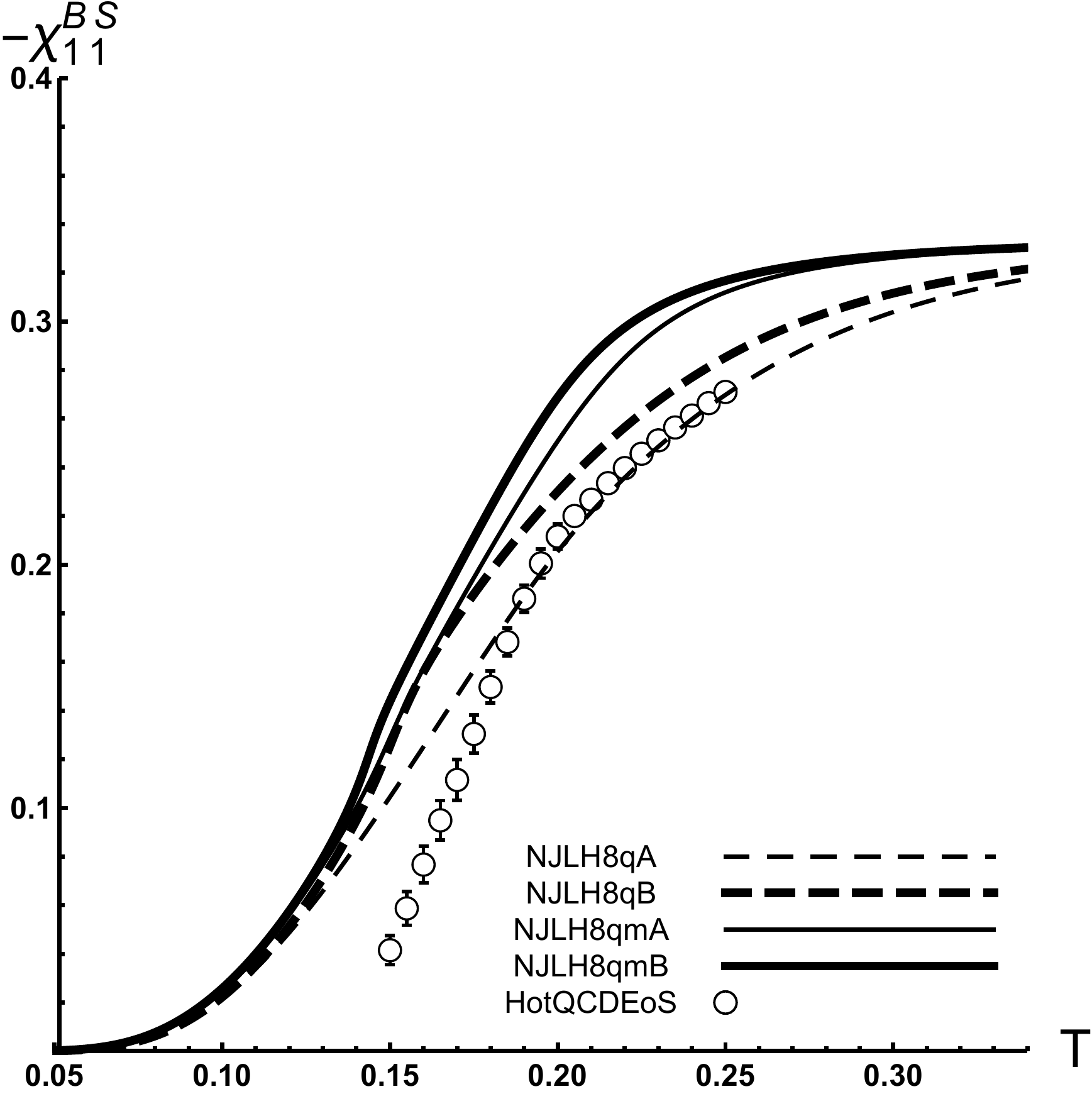}}
\subfigure[]{\label{Chi2QSELAPV}\includegraphics[width=0.32\textwidth]{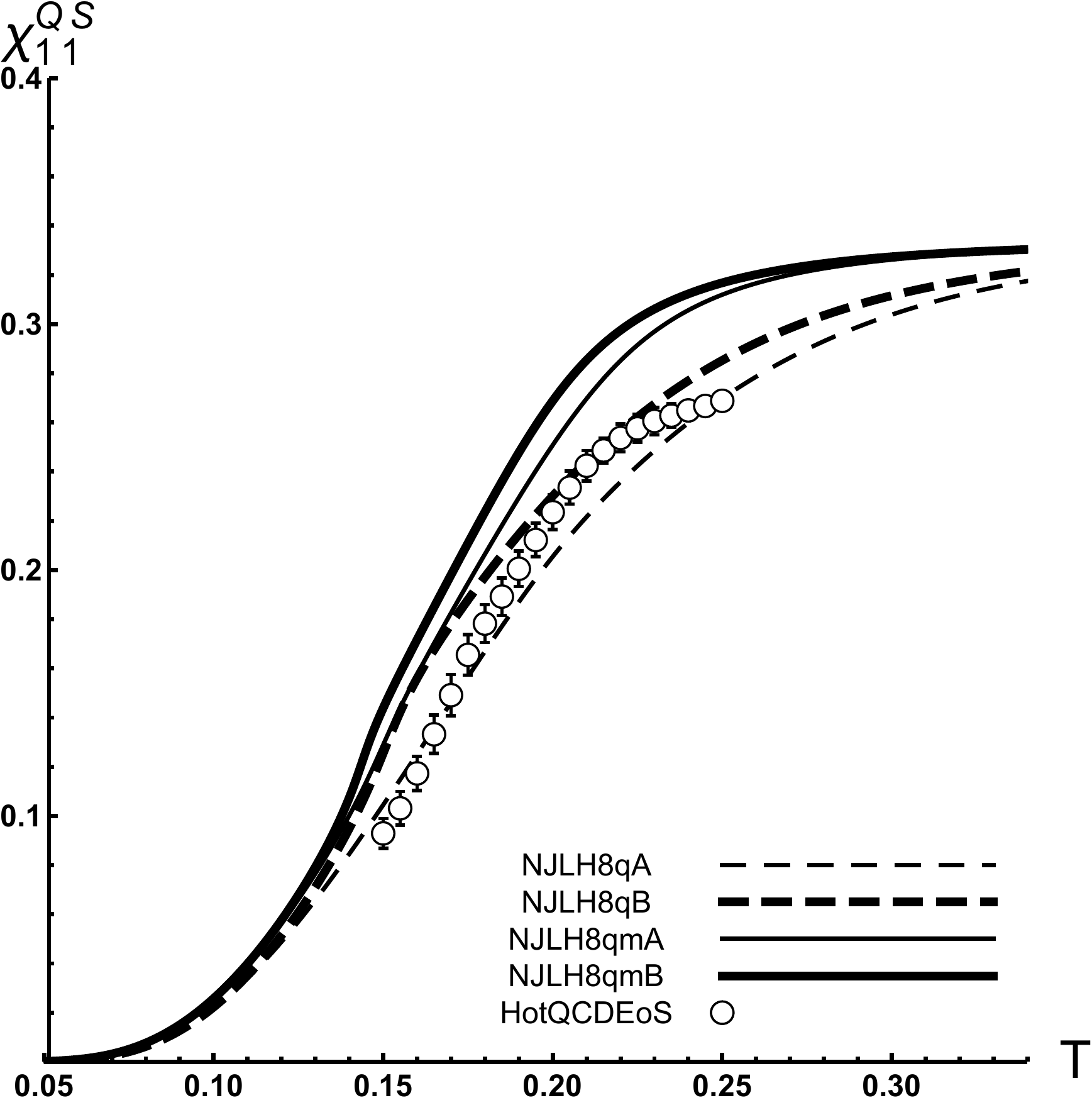}}
\caption{
Correlations of conserved charges as functions of temperature ($[T]=\mathrm{GeV}$) at vanishing chemical potential compared to lQCD results: in \ref{Chi2BQELAPV} correlation of baryonic number and electric charge ($\chi^{B\;Q}_{1\;1}$), in \ref{Chi2BSELAPV} baryonic number and strangeness ($\chi^{B\;S}_{1\;1}$) and in \ref{Chi2QSELAPV} electric charge and strangeness ($\chi^{Q\;S}_{1\;1}$). The markers correspond to continuum extrapolated lQCD results taken from \cite{Bazavov:2012jq}.
}
\label{CorrelationsELAPV}
\end{figure*}

In Fig. \ref{FluctuationsELAPV} the fluctuations of several conserved charges are shown. The main gross feature for the baryonic susceptibility $\chi_2^B$ in Fig. \ref{Chi2BELAPV} is that the slope improves significantly for the sets with strong $g_1$ coupling, in comparison with lQCD data. One sees further a  noticeable change of slope between the steep rise and  the flattening  of the curves for sets including ESB interactions, NJLH8qmA and NJLH8qmB. A hint of such behavior seems to be present in the  HotQCDEoS data as well. In a similar fashion do the slopes get improved in the fluctuations of strangeness $\chi_2^S$, shown in Fig. \ref{Chi2SELAPV} for the sets NJLH8qmA and NJLH8qmB. As opposed to these the slope is much steeper for these sets in the calculation of the electric charge fluctuations $\chi_2^Q$ displayed in Fig. \ref{Chi2QELAPV}, when compared to the lattice points. By expressing these fluctuations in terms of the quark number susceptibilities  %\cite{Borsanyi:2011sw} 
\begin{eqnarray}
\label{qnf}
&& \chi_2^B=\frac{1}{9}(\chi_2^u+\chi_2^d+\chi_2^s+2\chi_{11}^{us}+2\chi_{11}^{ds}+2\chi_{11}^{ud}) \nonumber \\
&& \chi_2^Q=\frac{1}{9}(4\chi_2^u+\chi_2^d+\chi_2^s-4\chi_{11}^{us}+2\chi_{11}^{ds}-4\chi_{11}^{ud}) \nonumber \\
&& \chi_2^S=\chi_2^s 
\end{eqnarray}
one sees that the weight of $\chi_2^u$ in $\chi_2^Q$ is larger than in $\chi_2^B$ by a factor 4. We have looked at the individual contributions within the model and found that the transition for $\chi_2^u$ occurs faster than for $\chi_2^s$, as expected, and that the slope increases when the ESB interactions are taken into account, while the crossed contributions vanish, reflecting the fact that the model has no gluonic degrees of freedom \cite{Koch:2005vg,Borsanyi:2011sw}, (the Polyakov loop introduces such a correlation, see section \ref{PNJLext} below). So it seems that the slope of $\chi_2^u$ dominates the scene in $\chi_2^Q$, due to the weighting factor, while the distribution of weights in the $\chi_2^B$  leads to the correct slope in comparison to lattice results. The $\chi_2^S$ provides for a clean probe of the strange quark number susceptibility, as this is the only contribution. For this case one sees that the lQCD slope is well reproduced with the ESB  model sets. 

Regarding the correlations displayed in Fig. \ref{CorrelationsELAPV}, the same effect seems to be at work for correlation of baryonic  and electric charges $\chi_{11}^{BQ}$, shown in Fig. \ref{Chi2BQELAPV}; it displays a too fast increase as compared to lQCD for the sets with ESB breaking. As opposed to this the correlations of baryonic and strangeness charges $\chi_{11}^{BS}$ in Fig. \ref {Chi2BSELAPV} show a slope in conformity with lQCD. The correlation of strangeness and electric charges also gets improved for the ESB sets, Fig. \ref{Chi2QSELAPV}. This can also be understood by looking at the dependence of these correlations on the quark number susceptibilities

\begin{eqnarray}
\label{correl}
&& \chi_{11}^{BQ}=\frac{1}{9}(2\chi_2^u-\chi_2^d-\chi_2^s+\chi_{11}^{ud}+\chi_{11}^{us}-2\chi_{11}^{ds}) \nonumber \\
&& \chi_{11}^{BS}=-\frac{1}{3}(\chi_2^s+\chi_{11}^{us}+\chi_{11}^{ds}) \nonumber \\
&& \chi_{11}^{QS}=\frac{1}{3}(\chi_2^s-2\chi_{11}^{us}+\chi_{11}^{ds}) 
\end{eqnarray}

Since the correlations $\chi_{11}^{us}$    %$i\ne j \subset u,d,s$
are smaller in magnitude in lQCD than the $\chi_2^i$ $(i=u,s)$  \cite{Borsanyi:2011sw} ($\chi_{11}^{ud}$ was shown to be small in the flavor $SU(2)$ case \cite{Allton:2005gk}), and vanish identically in the model, the only dependence is on $\chi_2^s$ for  $\chi_{11}^{BS},\chi_{11}^{QS}$. We have seen that the slope for $\chi_2^s$  reproduces well the corresponding lQCD slope, which explains the satisfactory behavior of the slopes of  $\chi_{11}^{BS},\chi_{11}^{QS}$ as well. The situation is different for $\chi_{11}^{BQ}$, which depends on $\chi_2^u$, for which a too steep slope occurred compared to lQCD.

\subsection{PNJL extension}
\label{PNJLext}

The impact of coupling the quark degrees of freedom to the gluonic sector, using the PNJL model extension with two types of potentials is discussed in this section. 

The gross feature is a systematic shift of all the curves describing the observables of the last subsection to higher temperatures, which is an important effect in bringing most of the observables related with fluctuations and correlated charges closer to the lQCD curves. 

However the effect on the velocity of sound and the trace of the energy momentum tensor depends strongly on the type of Polyakov loop potential used. % is overpowering, superseding all the fine nuances discussed previously in relation with %ESB terms.
Let us discuss first these observables.

\begin{figure*}[htb]
\center
\subfigure[]{\label{Cs2PELAPVLogTc200}\includegraphics[width=0.32\textwidth]{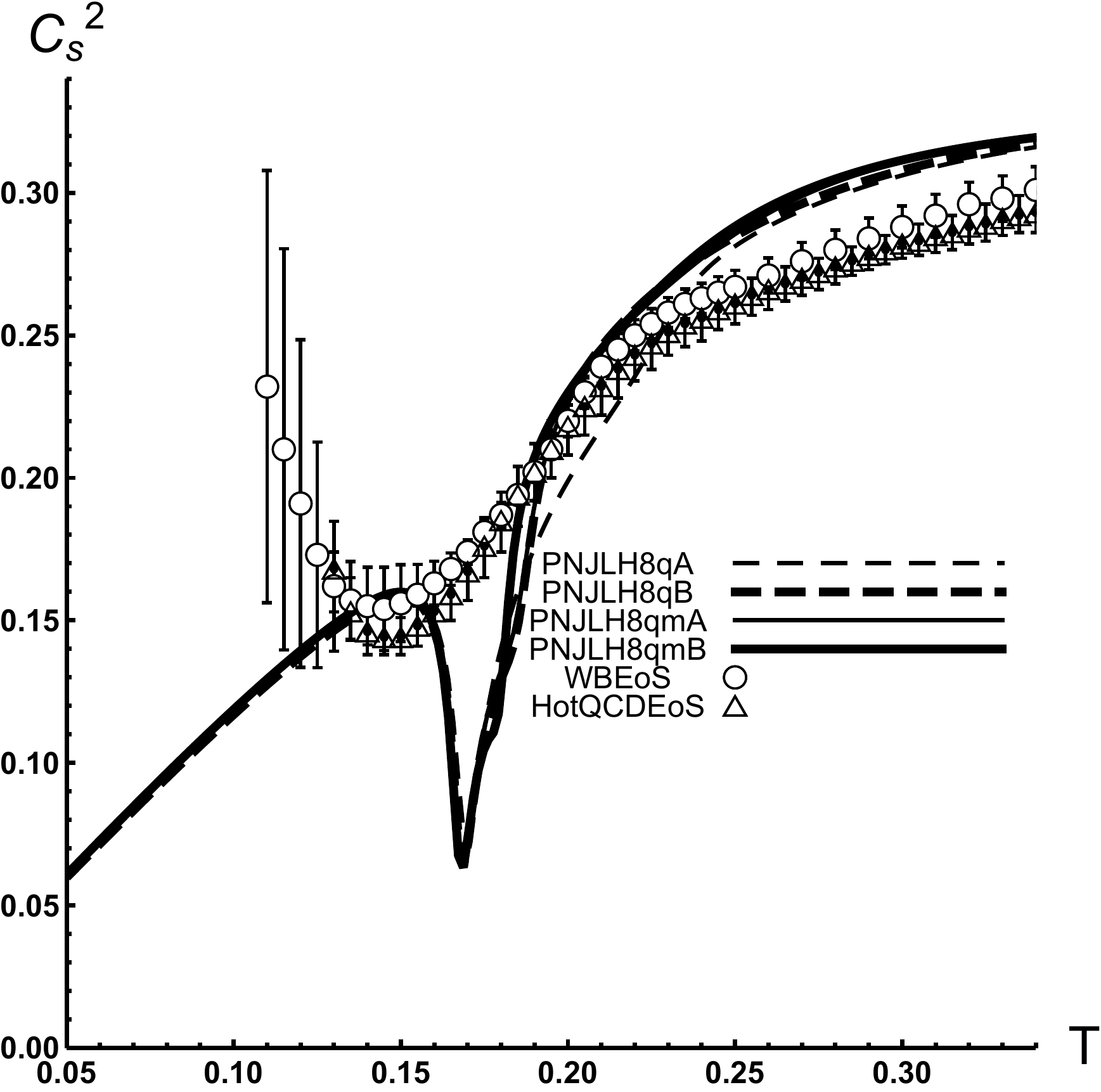}}
\subfigure[]{\label{TmumuPELAPVLogTc200}\includegraphics[width=0.32\textwidth]{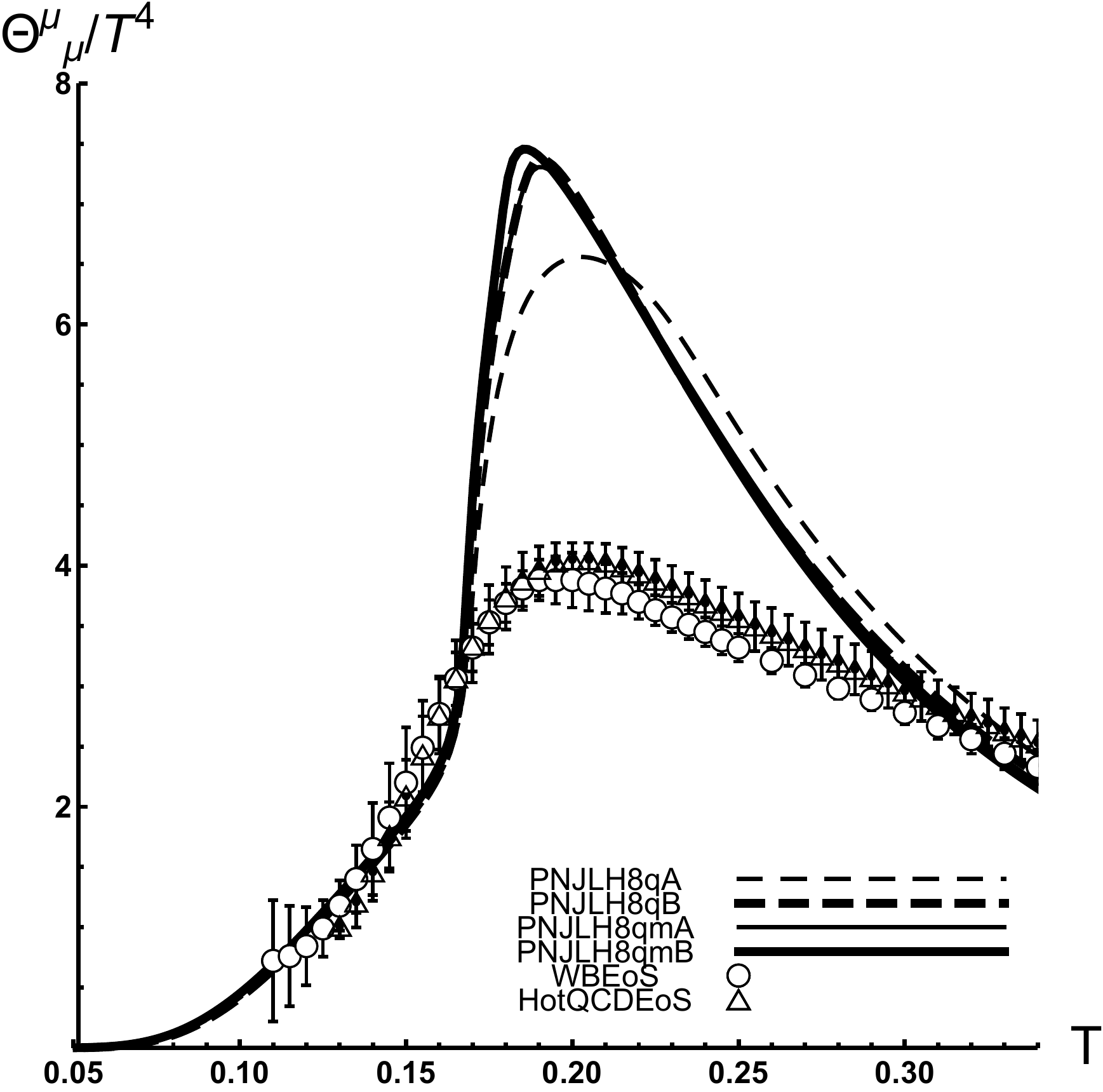}}
\caption{
In \ref{Cs2PELAPVLogTc200} the model Polyakov loop extension with potential ${\cal U}_I$ in Eq. \ref{UPI} is shown for the squared speed of sound as a function of temperature ($[T]=\mathrm{GeV}$) at vanishing chemical potential as obtained using the parameter sets from Table \ref{ParameterSets}. In the legend  a "P" has been attached at the beginning of each parameter set, meaning that the Polyakov loop extension has been applied, PNJLH8qA and PNJLH8qB correspond to the sets without the ESB terms, PNJLH8qmA, PNJLH8qmB include the ESB interactions.The markers, labeled as WBEoS and HotQCDEoS, correspond to continuum extrapolated lQCD results taken respectively from \cite{Borsanyi:2013bia} and \cite{Bazavov:2014pvz}.
In \ref{TmumuPELAPVLogTc200} is shown the energy-momentum trace anomaly for the same Polyakov loop potential.
}
\label{Cs2TrAnomPELAPVLogTc200}
\end{figure*}

\begin{figure*}[htb]
\center
\subfigure[]{
\label{Cs2PELAPVExpKLogBhatTc175}
\includegraphics[width=0.32\textwidth]{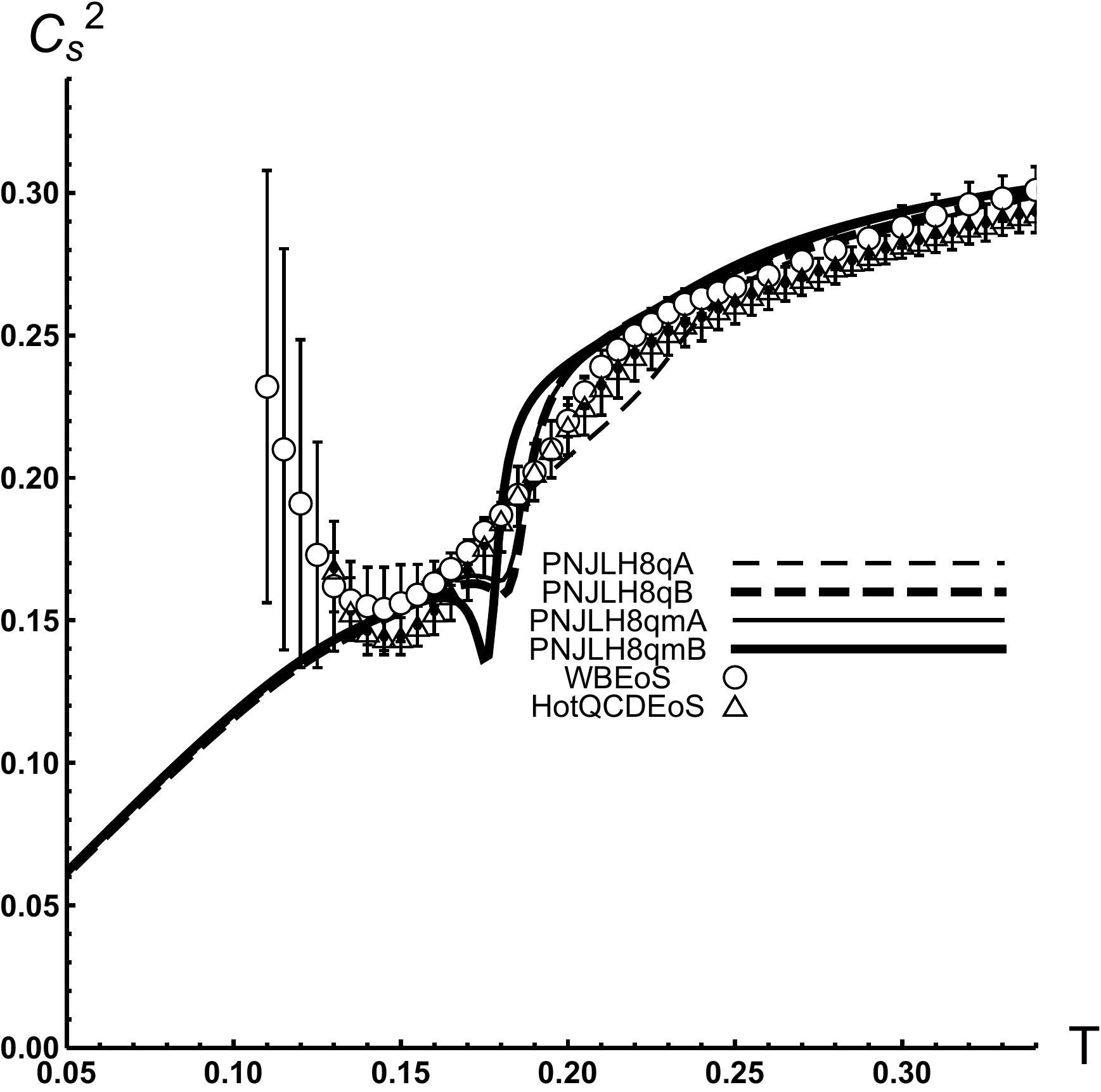}}
\subfigure[]{
\label{TmumuPELAPVExpKLogBhatTc175}
\includegraphics[width=0.32\textwidth]{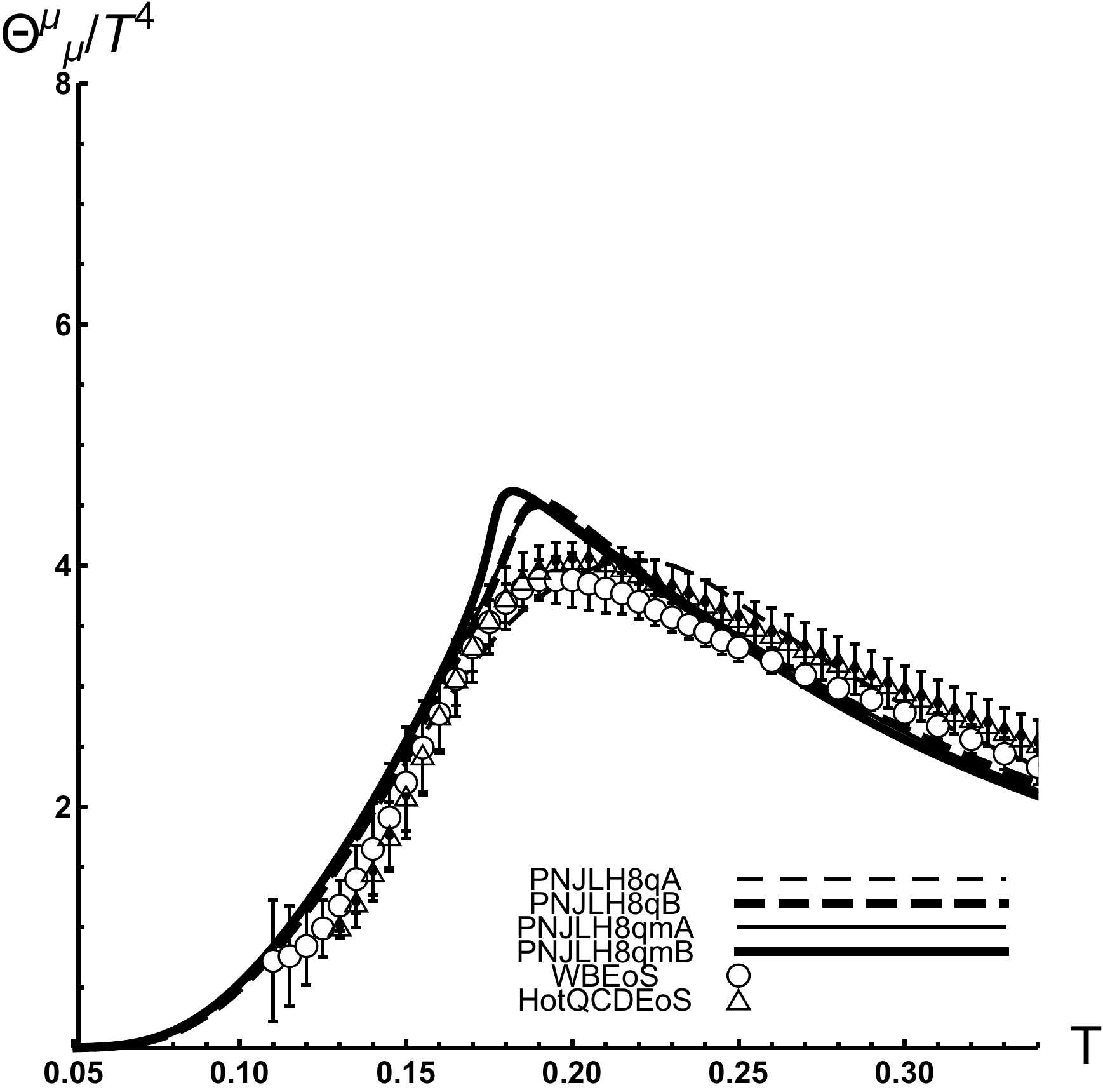}}
\caption{ In \ref{Cs2PELAPVExpKLogBhatTc175} and  \ref{TmumuPELAPVExpKLogBhatTc175}  the same observables are shown as in  \ref{Cs2PELAPVLogTc200} and \ref{TmumuPELAPVLogTc200}, but obtained with the Polyakov loop potential  ${\cal U}_{II}$ in Eq. \ref{UPII}. The markers, labeled as WBEoS and HotQCDEoS, correspond to continuum extrapolated lQCD results taken respectively from \cite{Borsanyi:2013bia} and \cite{Bazavov:2014pvz}. 
}
\label{Cs2TrAnomPELAPVExpKLogBhatTc175}
\end{figure*}

In Fig. \ref{Cs2PELAPVLogTc200}  the velocity of sound is displayed,  calculated with the Polyakov loop potential ${\cal U}_I$ in Eq. \ref{UPI},  showing that independently of the NJL parameter sets of Table \ref{ParameterSets} considered, a too deep relative minimum for the velocity of sound occurs, about a factor 2.5 smaller in magnitude, in comparison with lQCD. This result supersedes all the nuances discussed previously in relation with ESB terms. In Fig. \ref{TmumuPELAPVLogTc200} one sees that the peak of the trace of the energy momentum tensor is roughly twice the value of the lQCD one. These dominating characteristics are also present in the polylogarithmic variant of the Polyakov loop potential in \cite{Stiele:2016cfs}, which are therefore not shown. 

Contrary to this, the potential  ${\cal U}_{II}$ in Eq. \ref{UPII} discriminates between the different NJL sets.   Minima occur for the sets with ESB interactions (and large $g_1$ coupling without ESB) see Fig.  \ref{Cs2PELAPVExpKLogBhatTc175}, two shallow minima, and a more pronounced one for the ESB set with stronger $g_1$ coupling;  the set  PNJLH8qA  without ESB terms corresponding to weak $g_1$ coupling does not display a minimum in the velocity of sound, as also verified in \cite{Bhattacharyya:2017gwt} using ${\cal U}_{II}$, and as it was the case without the Polyakov loop extension, see Fig. \ref{Cs2ELAPV}. 

\begin{figure*}[htb]
\center
\subfigure[]{\label{PPELAPVExpKLogBhatTc175}\includegraphics[width=0.32\textwidth]{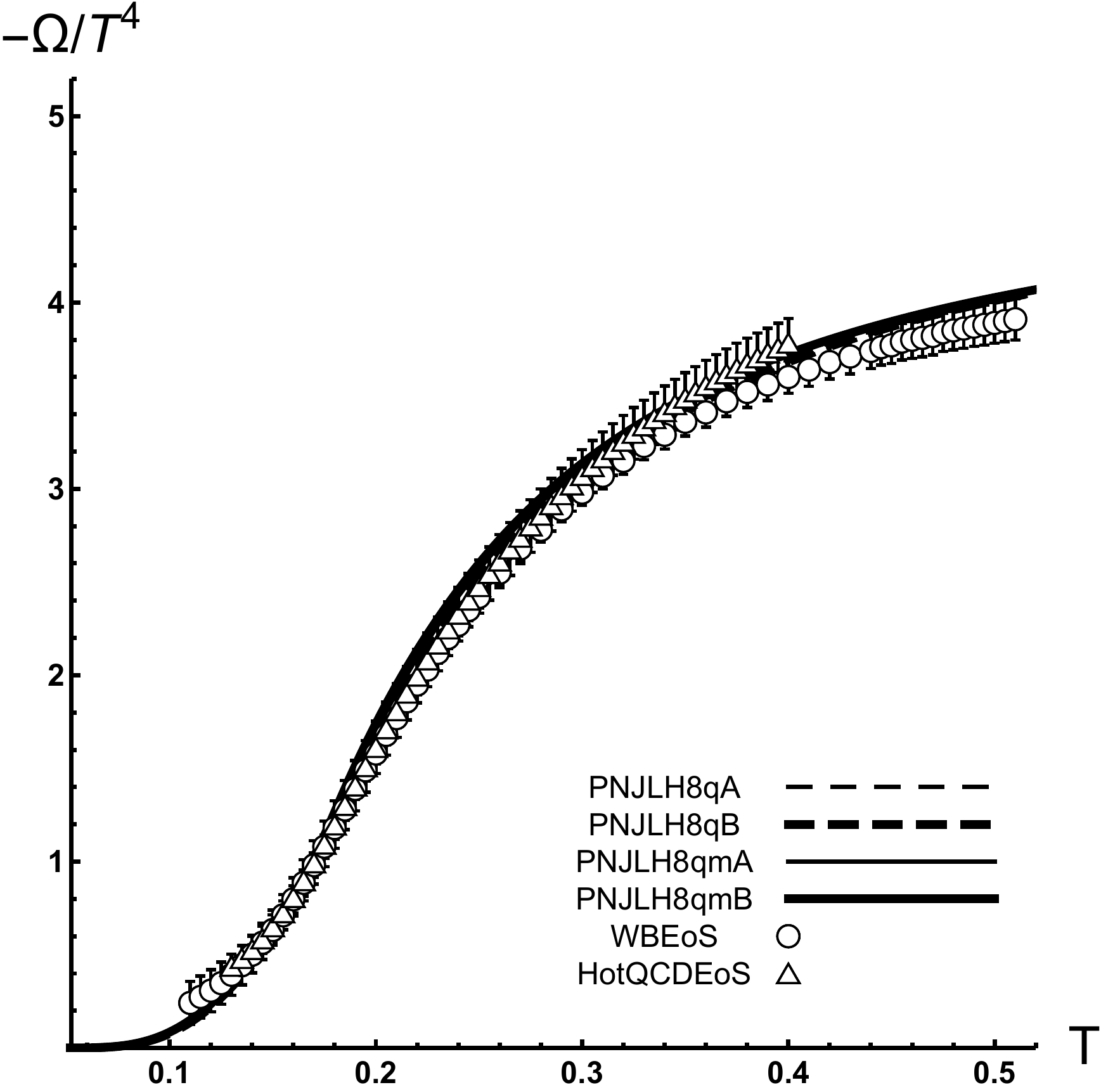}}
\subfigure[]{\label{epsPELAPVExpKLogBhatTc175}\includegraphics[width=0.32\textwidth]{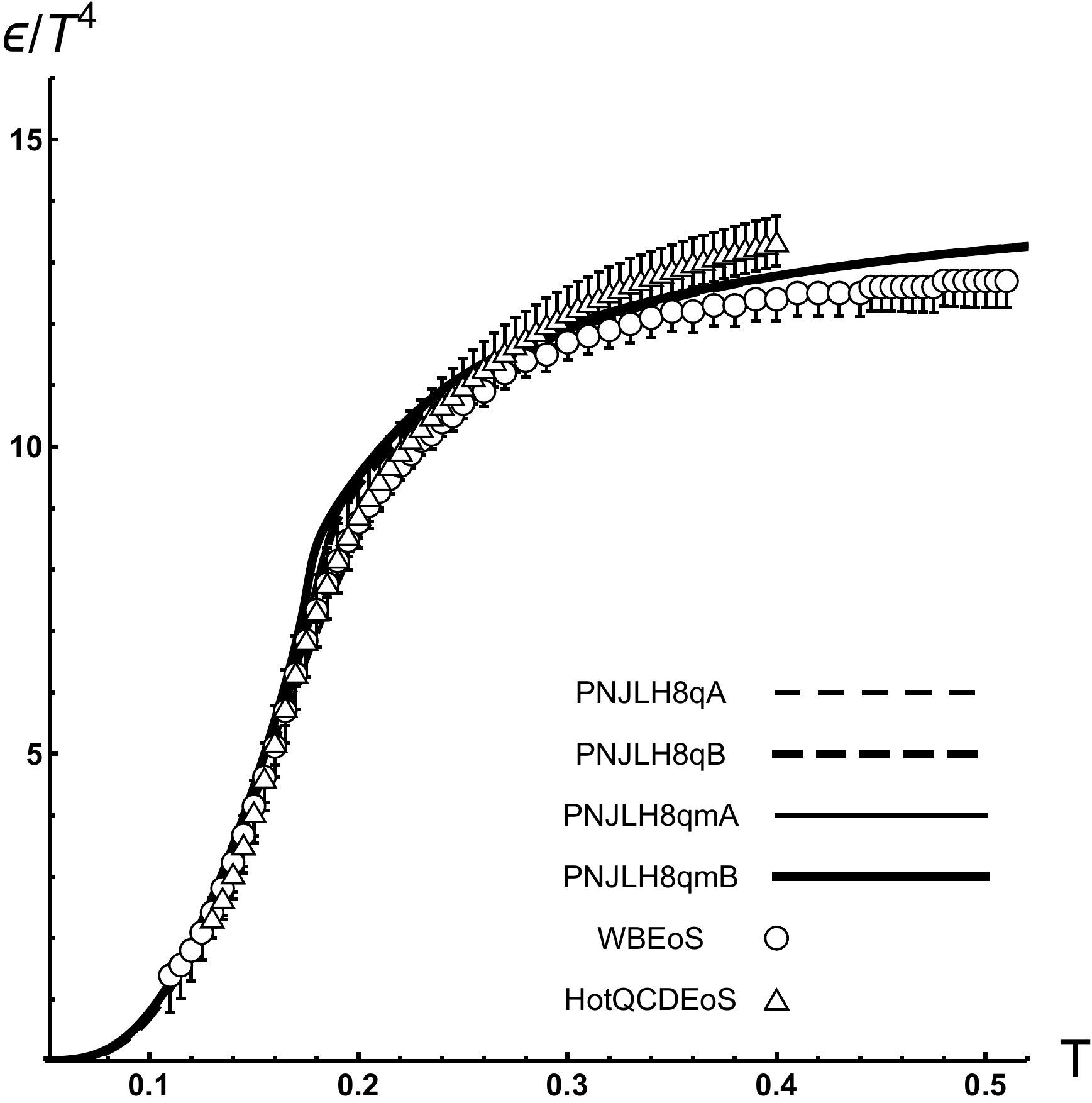}}\\
\subfigure[]{\label{SPELAPVExpKLogBhatTc175}\includegraphics[width=0.32\textwidth]{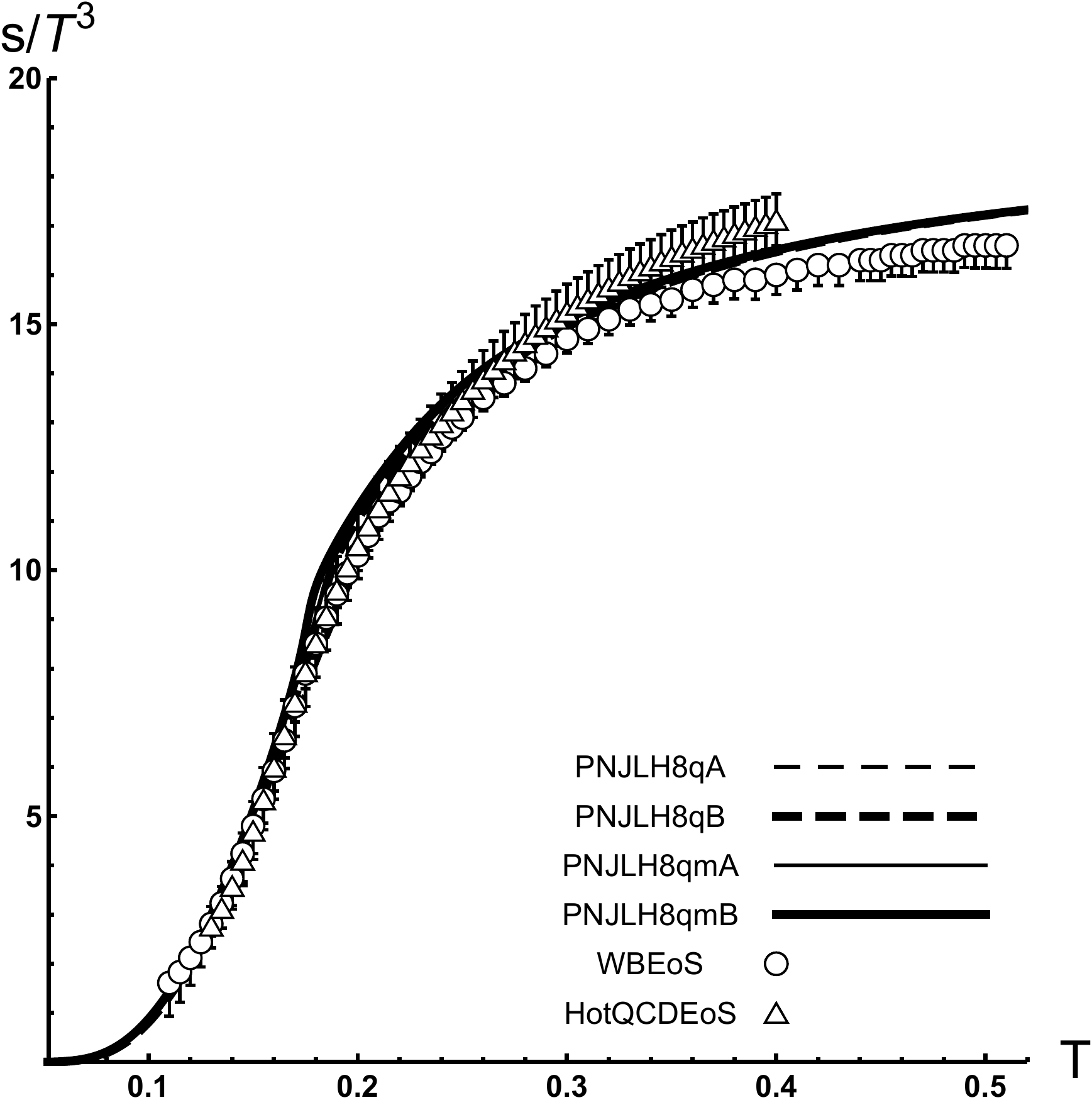}}
\subfigure[]{\label{CVPELAPVExpKLogBhatTc175}\includegraphics[width=0.32\textwidth]{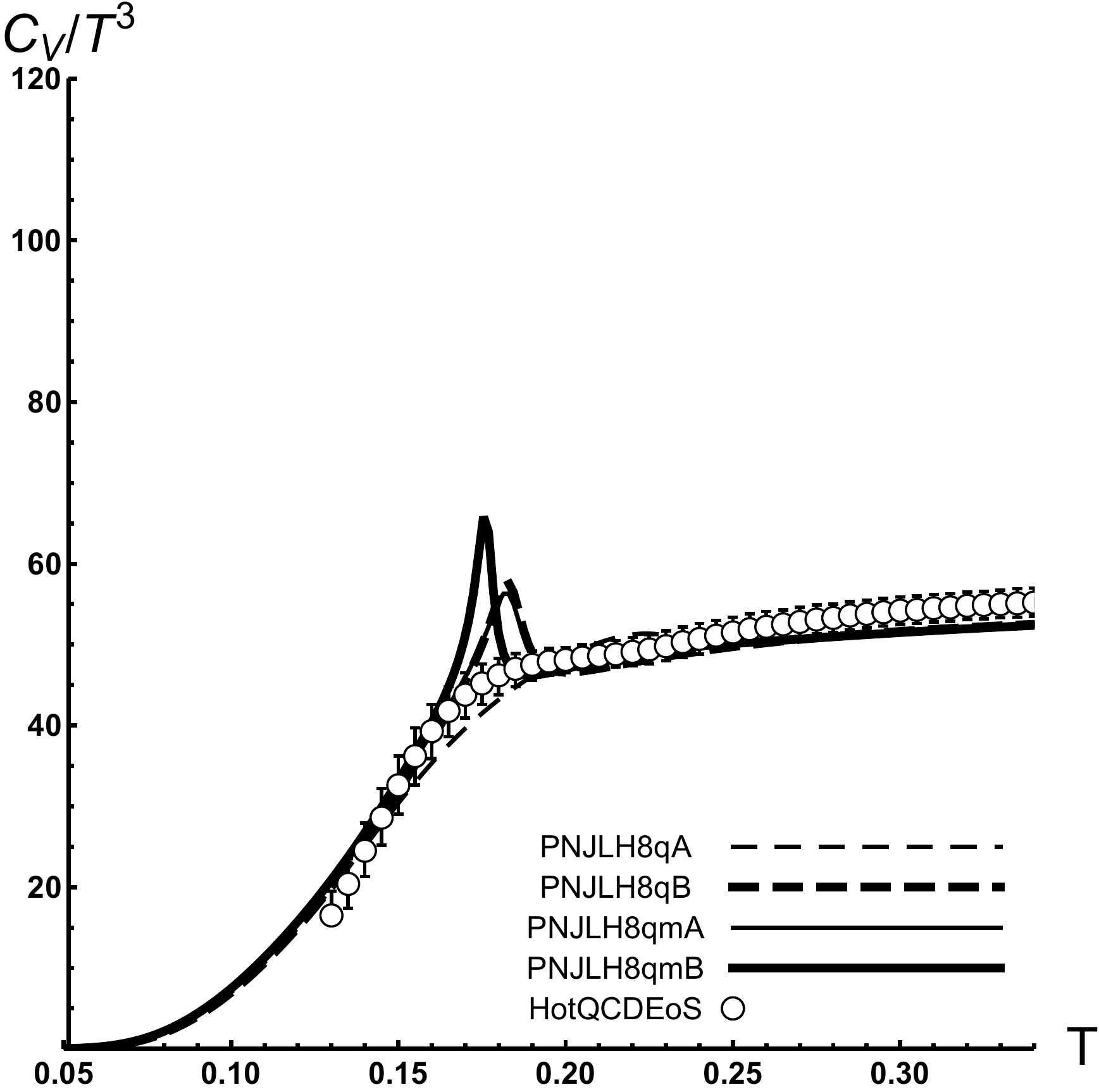}}
\caption{
Pressure \ref{PPELAPVExpKLogBhatTc175}, energy density \ref{epsPELAPVExpKLogBhatTc175}, entropy \ref{SPELAPVExpKLogBhatTc175} and specific heat \ref{CVPELAPVExpKLogBhatTc175} (divided by the corresponding powers of temperature as to render them dimensionless, $P/T^4=-\Omega/T^4$, $\epsilon/T^4$, $s/T^3$ and $C_V/T^3$, respectively) as functions of temperature ( $[T]=\mathrm{GeV}$). The markers, labeled as WBEoS and HotQCDEoS, correspond to continuum extrapolated lQCD results taken respectively from \cite{Borsanyi:2013bia} and \cite{Bazavov:2014pvz}.
}
\label{PepsSCVPELAPVExpKLogBhatTc175}
\end{figure*}

Regarding the trace of the energy momentum tensor it turns out to be fairly well represented in comparison with the lQCD results using  ${\cal U}_{II}$, see Fig. \ref{TmumuPELAPVExpKLogBhatTc175}. The individual thermodynamical quantities contributing to Fig. \ref{TmumuPELAPVExpKLogBhatTc175}, $\epsilon$ and $P$, as well as their derivatives with respect to the temperature $T$, $C_V$ and $s$ are depicted for the case of the potential ${\cal U}_{II}$ in Figs \ref{PepsSCVPELAPVExpKLogBhatTc175}. Overall, the correspondence of the presented quantities with lQCD is quite satisfactory (note that the inclusion of the extra degrees of freedom enables the correct asymptotic behavior for $P$, $\epsilon$ and $s$). The slight change of slope in $\epsilon$ around $T=0.18$, compared to lQCD, results in a visible peak in $C_V$. For $C_V$ and ${\Theta^\mu}_\mu$ the quark interaction parameter set which presents the best fit is NJLH8qA. For the other quantities the difference between quark interaction parameter sets (mainly in the transition region) is too small for the purpose of comparison with lQCD results. 
We omit the corresponding figures for the choice of ${\cal U}_{I}$  since the calculations resulted in large deviations from the respective lQCD data, as one would expect from Fig. \ref{TmumuPELAPVLogTc200}, and turn out not to be very instructive. 

\begin{figure*}[htb]
\center
\subfigure[]{\label{Chi2BPELAPVLogTc200}\includegraphics[width=0.32\textwidth]{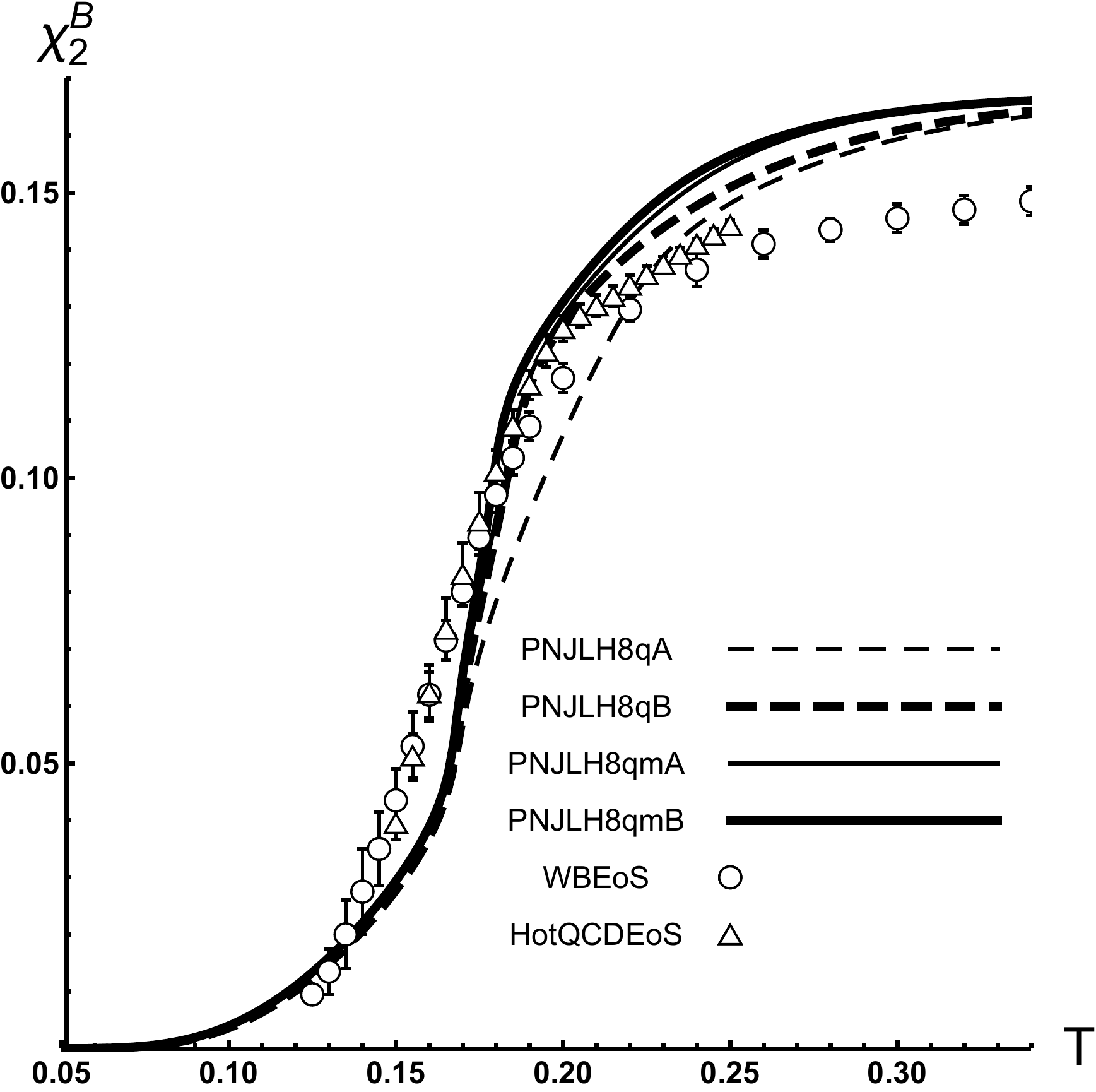}}
\subfigure[]{\label{Chi2QPELAPVLogTc200}\includegraphics[width=0.32\textwidth]{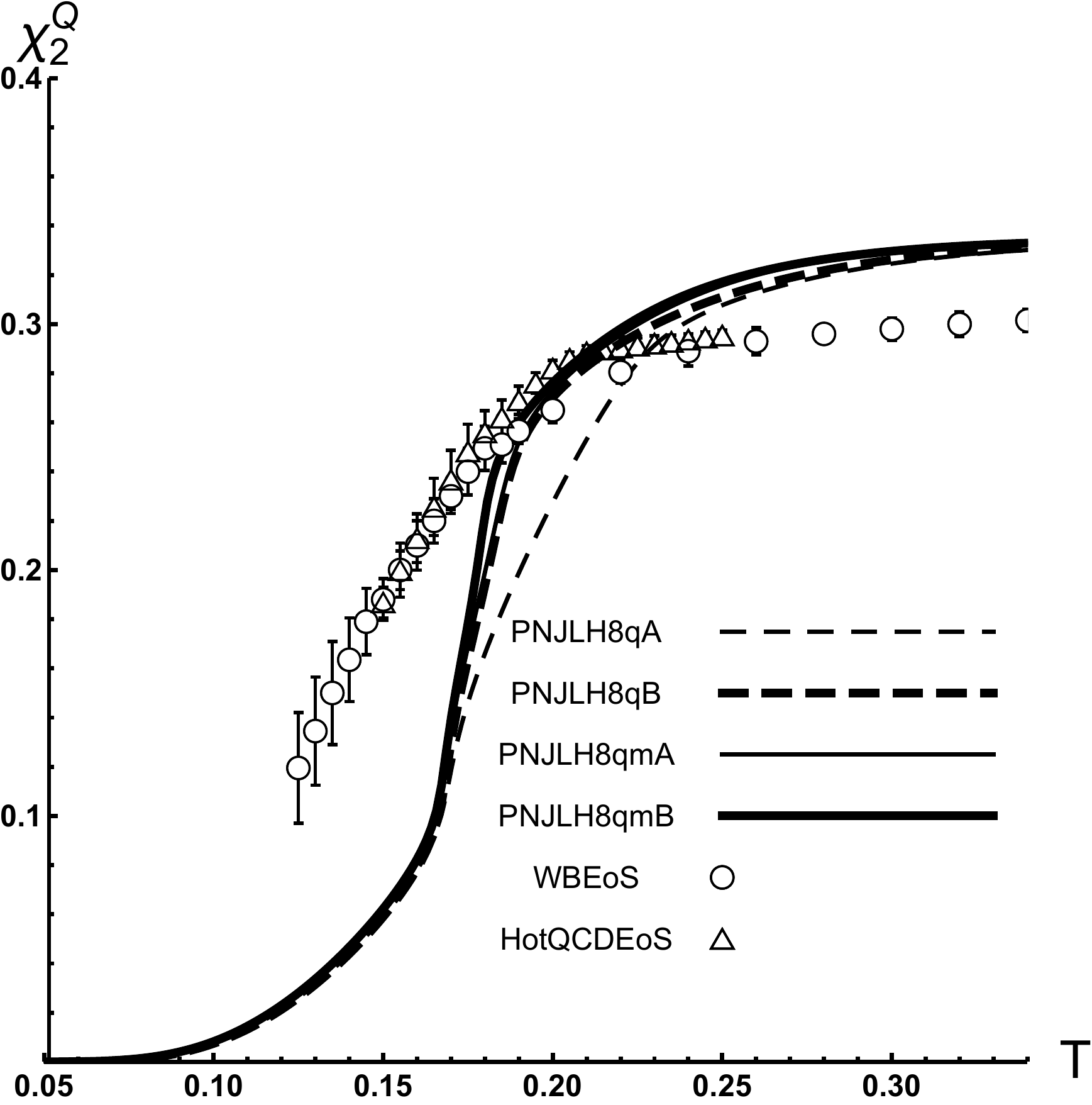}}
\subfigure[]{\label{Chi2SPELAPVLogTc200}\includegraphics[width=0.32\textwidth]{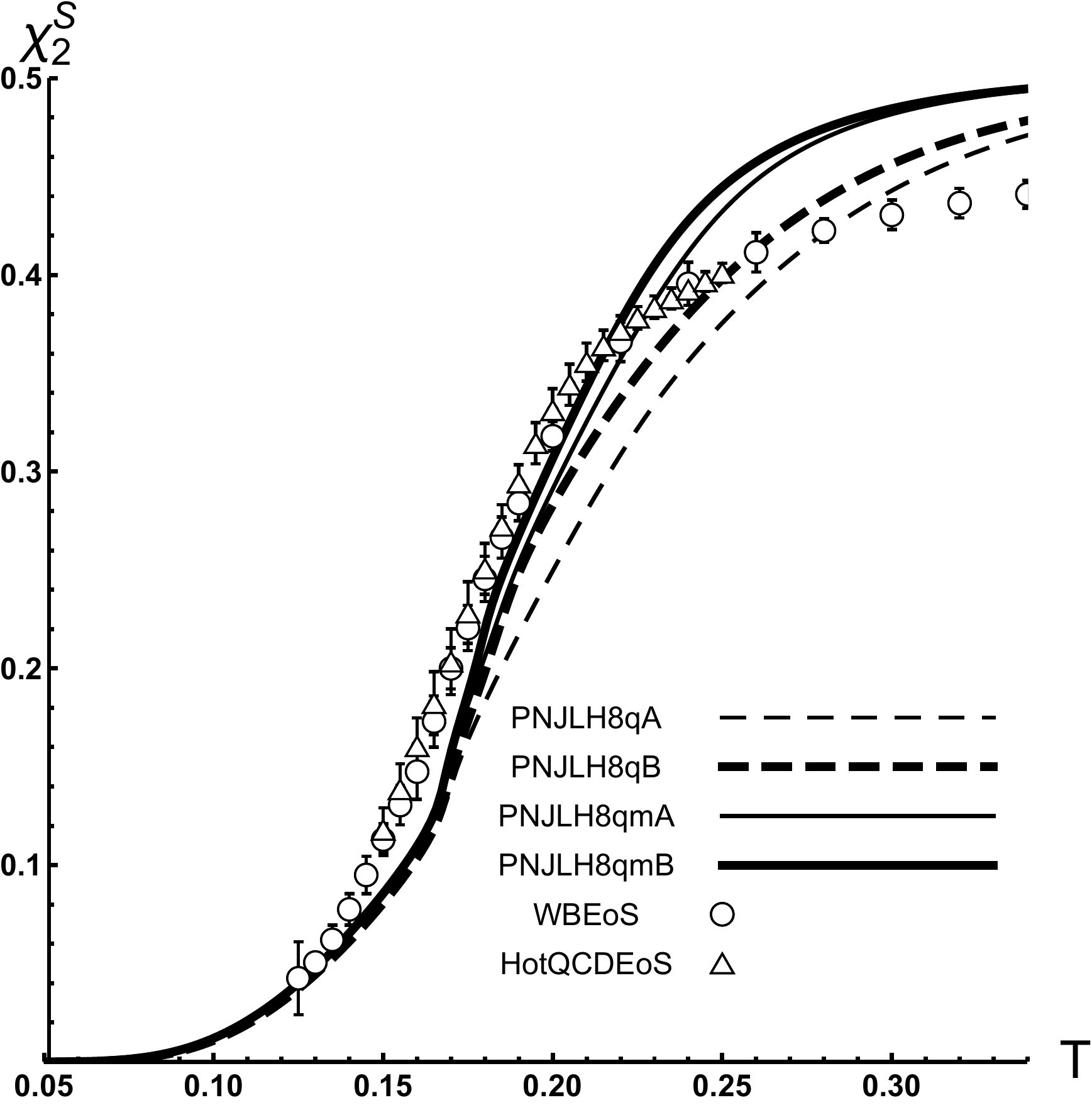}}
\caption{
Polyakov loop extension, using ${\cal U}_I$ in Eq. \ref{UPI}, for the fluctuations of conserved charges as functions of temperature ($[T]=\mathrm{GeV}$) at vanishing chemical potential, compared to lQCD results: in \ref{Chi2BPELAPVLogTc200} fluctuation of baryonic number ($\chi^B_2$), in \ref{Chi2QPELAPVLogTc200} electric charge ($\chi^Q_2$) and in \ref{Chi2SPELAPVLogTc200} strangeness ($\chi^S_2$). Same notation is used as in \ref{Cs2PELAPVLogTc200} for the lines.
The markers, labeled as WBEoS and HotQCDEoS, correspond to continuum extrapolated lQCD results taken respectively from \cite{Borsanyi:2011sw} and \cite{Bazavov:2012jq}. 
}
\label{FluctuationsPELAPVLogTc200}
\end{figure*}

\begin{figure*}[htb]
\center
\subfigure[]{
\label{Chi2BPELAPVExpKLogBhatTc175}
\includegraphics[width=0.32\textwidth]{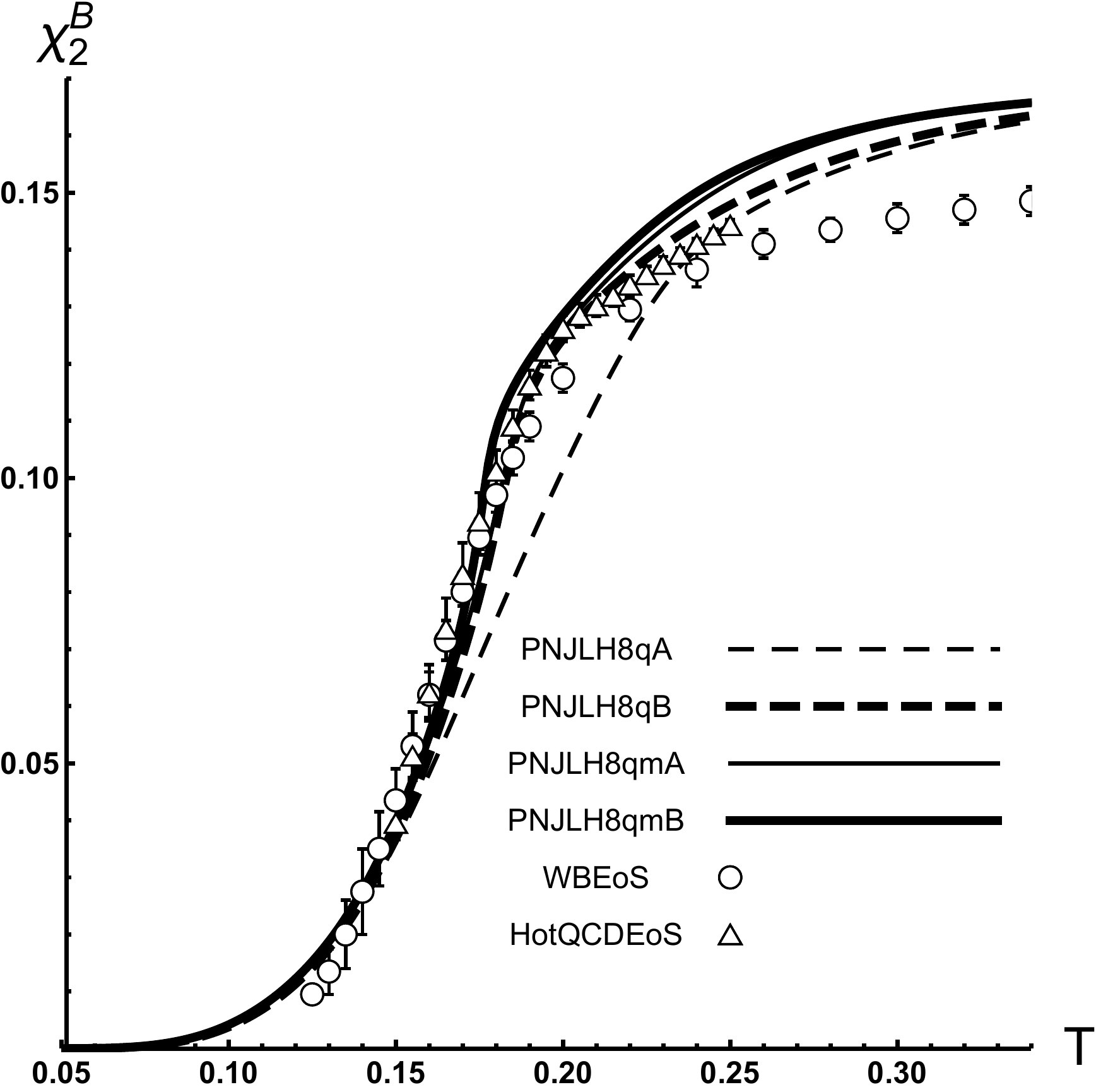}}
\subfigure[]{
\label{Chi2QPELAPVExpKLogBhatTc175}
\includegraphics[width=0.32\textwidth]{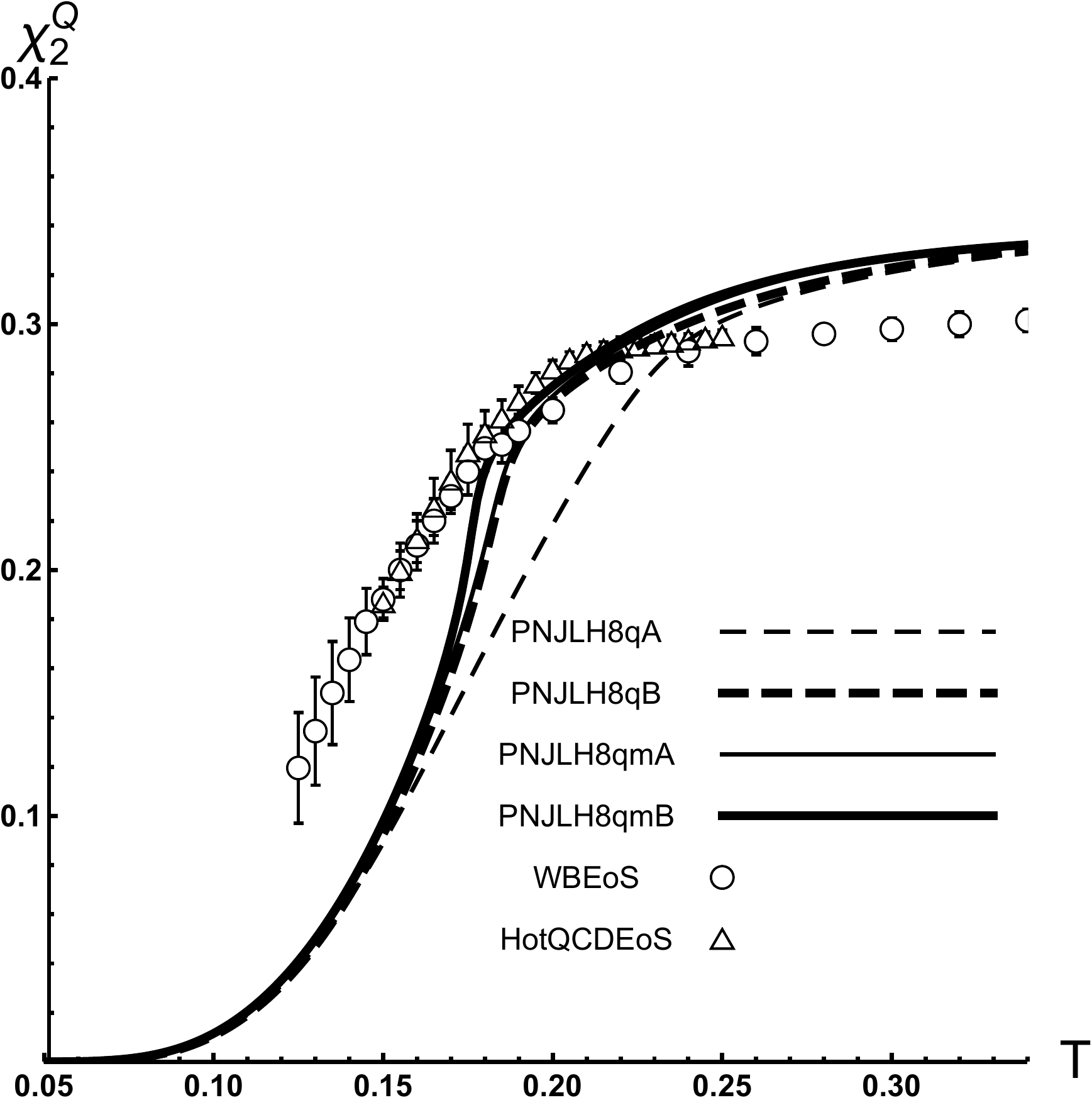}}
\subfigure[]{
\label{Chi2SPELAPVExpKLogBhatTc175}
\includegraphics[width=0.32\textwidth]{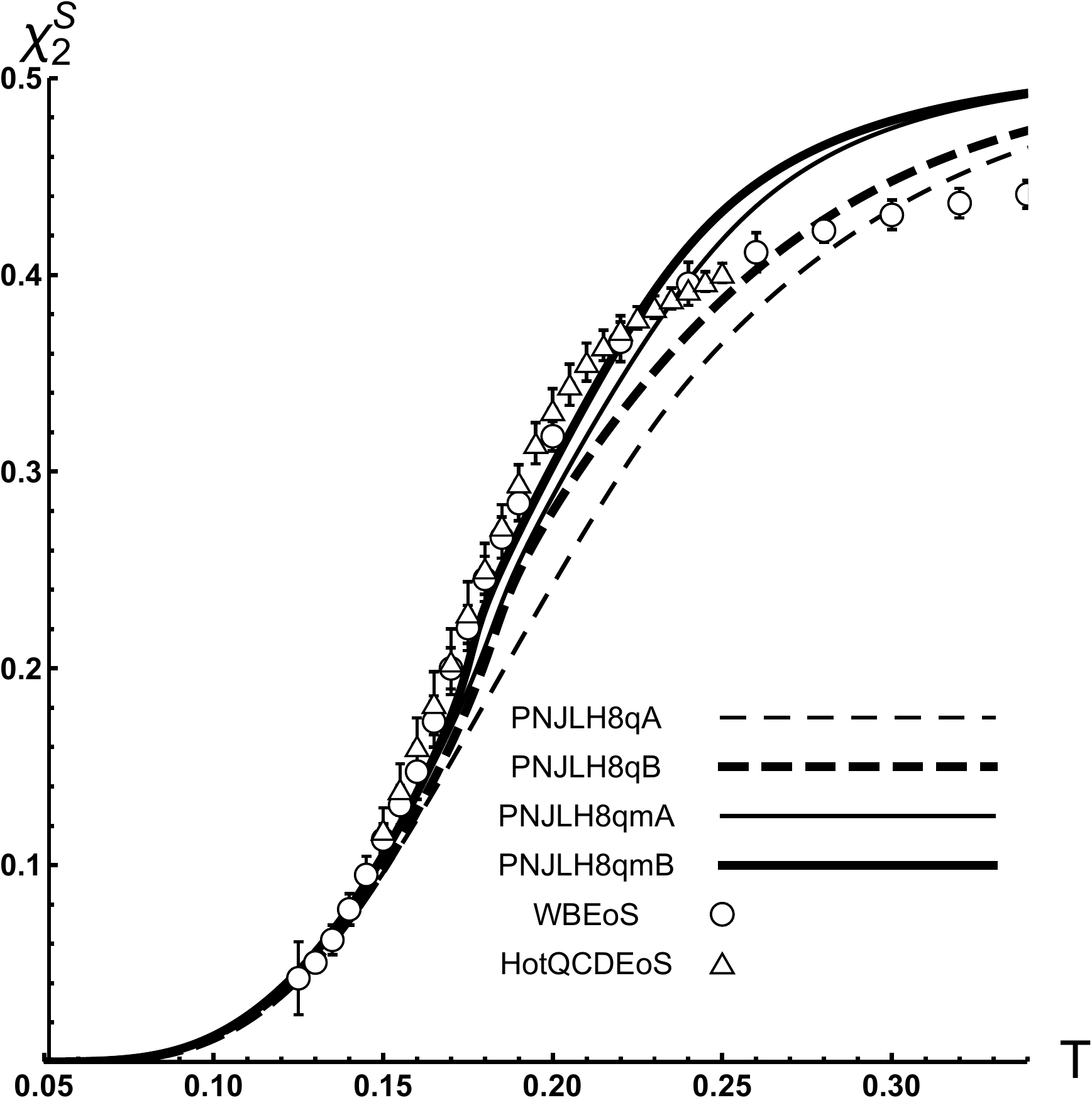}}
\caption{
Polyakov loop extension, using ${\cal U}_{II}$ in Eq. \ref{UPII} for the same observables as in \ref{Chi2BPELAPVLogTc200},\ref{Chi2QPELAPVLogTc200},\ref{Chi2SPELAPVLogTc200}.  
The markers, labeled as WBEoS and HotQCDEoS, correspond to continuum extrapolated lQCD results taken respectively from \cite{Borsanyi:2011sw} and \cite{Bazavov:2012jq}. 
}
\label{FluctuationsPELAPVExpKLogBhatTc175}
\end{figure*}

\begin{figure*}[htb]
\center
\subfigure[]{\label{Chi2BQPELAPVLogTc200}\includegraphics[width=0.32\textwidth]{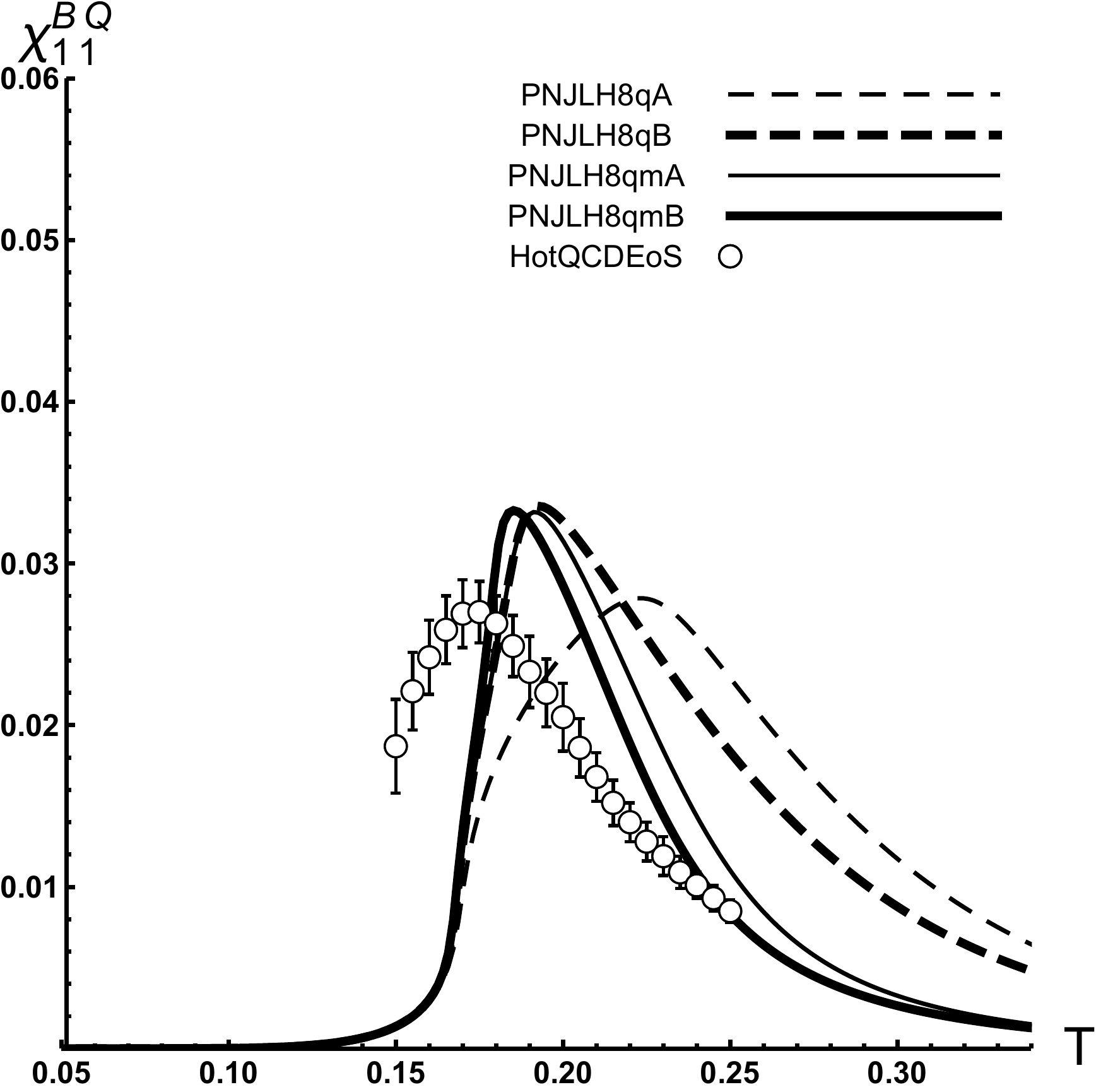}}
\subfigure[]{\label{Chi2BSPELAPVLogTc200}\includegraphics[width=0.32\textwidth]{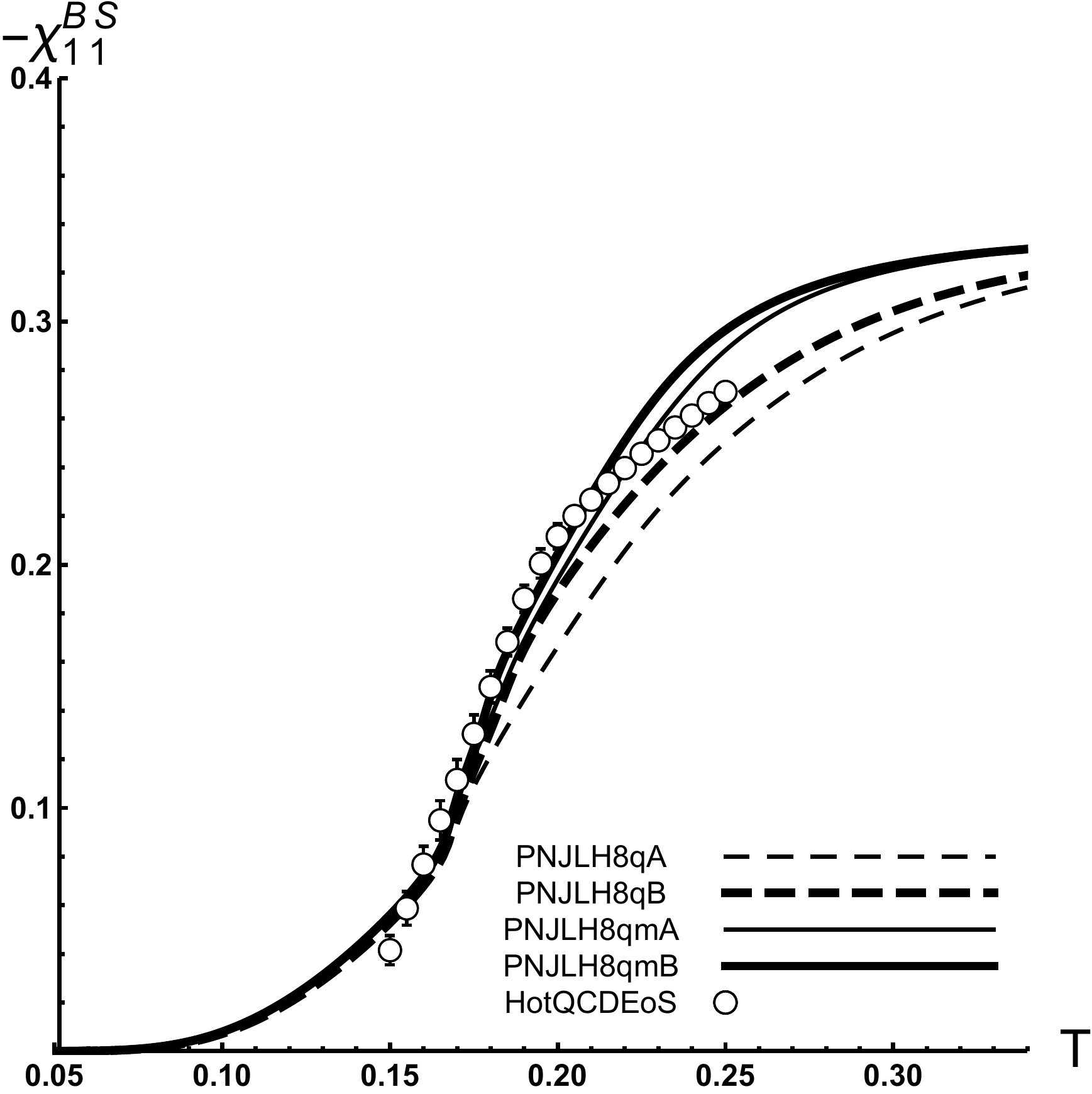}}
\subfigure[]{\label{Chi2QSPELAPVLogTc200}\includegraphics[width=0.32\textwidth]{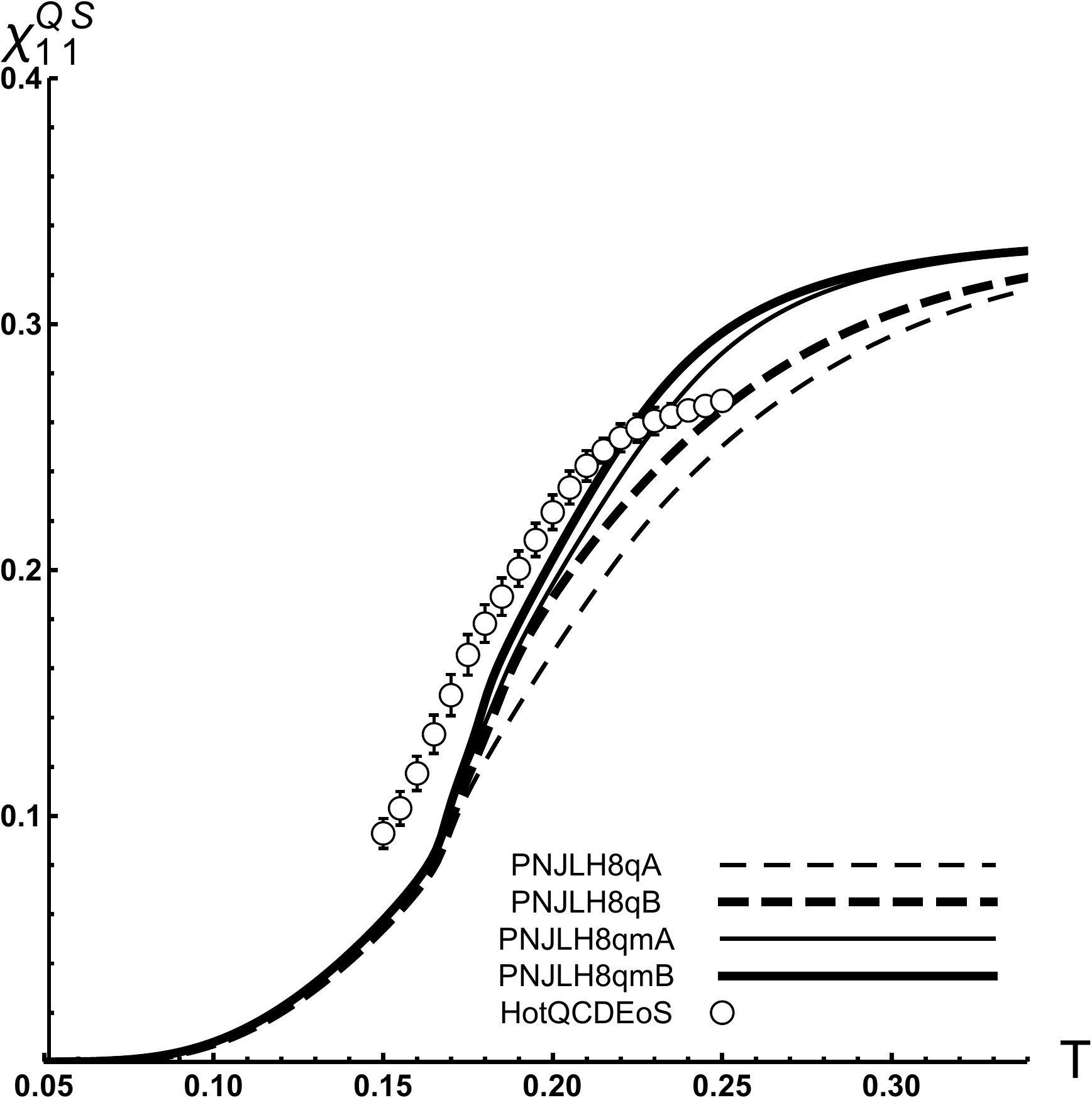}}
\caption{
Polyakov loop extension using ${\cal U}_I$ in Eq. \ref{UPI} for the correlations of conserved charges as functions of temperature ($[T]=\mathrm{GeV}$) at vanishing chemical potential compared to lQCD results: in \ref{Chi2BQPELAPVLogTc200} correlation of baryonic number and electric charge ($\chi^{B\;Q}_{1\;1}$), in \ref{Chi2BSPELAPVLogTc200} baryonic number and strangeness ($\chi^{B\;S}_{1\;1}$) and in \ref{Chi2QSPELAPVLogTc200} electric charge and strangeness ($\chi^{Q\;S}_{1\;1}$). Same notation for lines as before. The markers correspond to continuum extrapolated lQCD results taken from \cite{Bazavov:2012jq}}.
%PNJL LOG TC=200MeV!!!!!!!!!!!!!!}
\label{CorrelationsPELAPVLogTc200}
\end{figure*}

\begin{figure*}[htb]
\center
\subfigure[]{
\label{Chi2BQPELAPVExpKLogBhatTc175}
\includegraphics[width=0.32\textwidth]{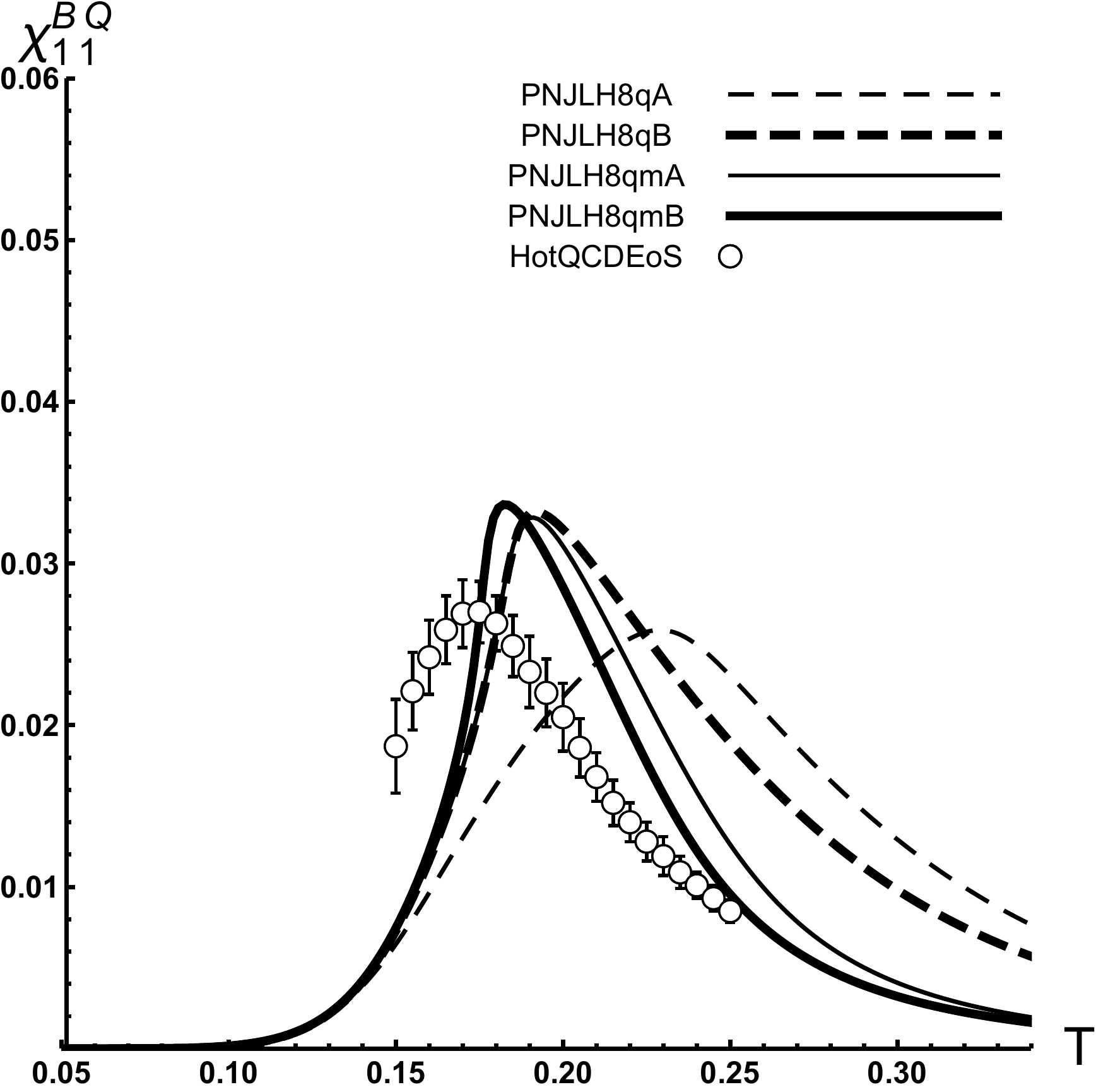}}
\subfigure[]{
\label{Chi2BSPELAPVExpKLogBhatTc175}
\includegraphics[width=0.32\textwidth]{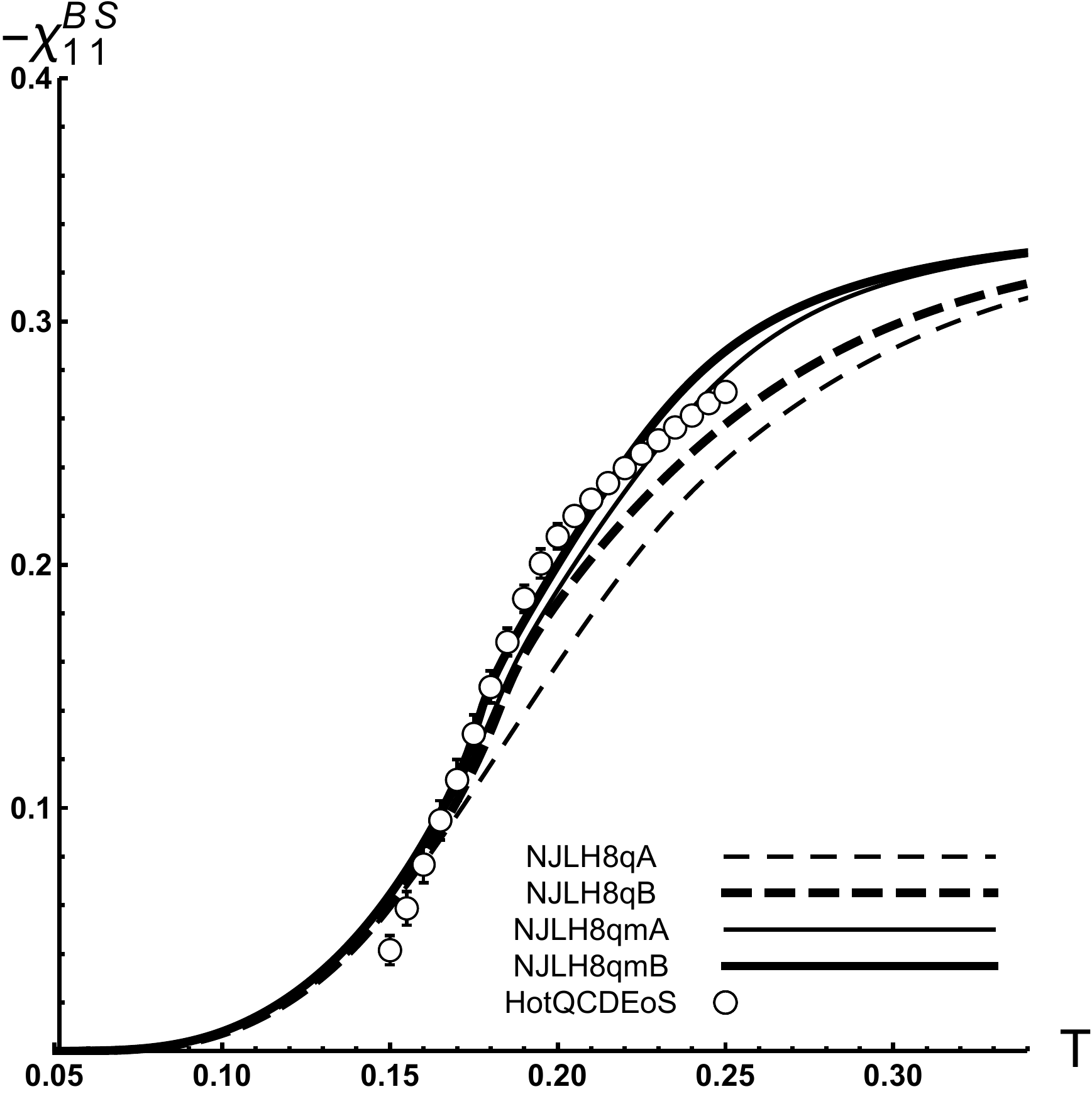}}
\subfigure[]{
\label{Chi2QSPELAPVExpKLogBhatTc175}
\includegraphics[width=0.32\textwidth]{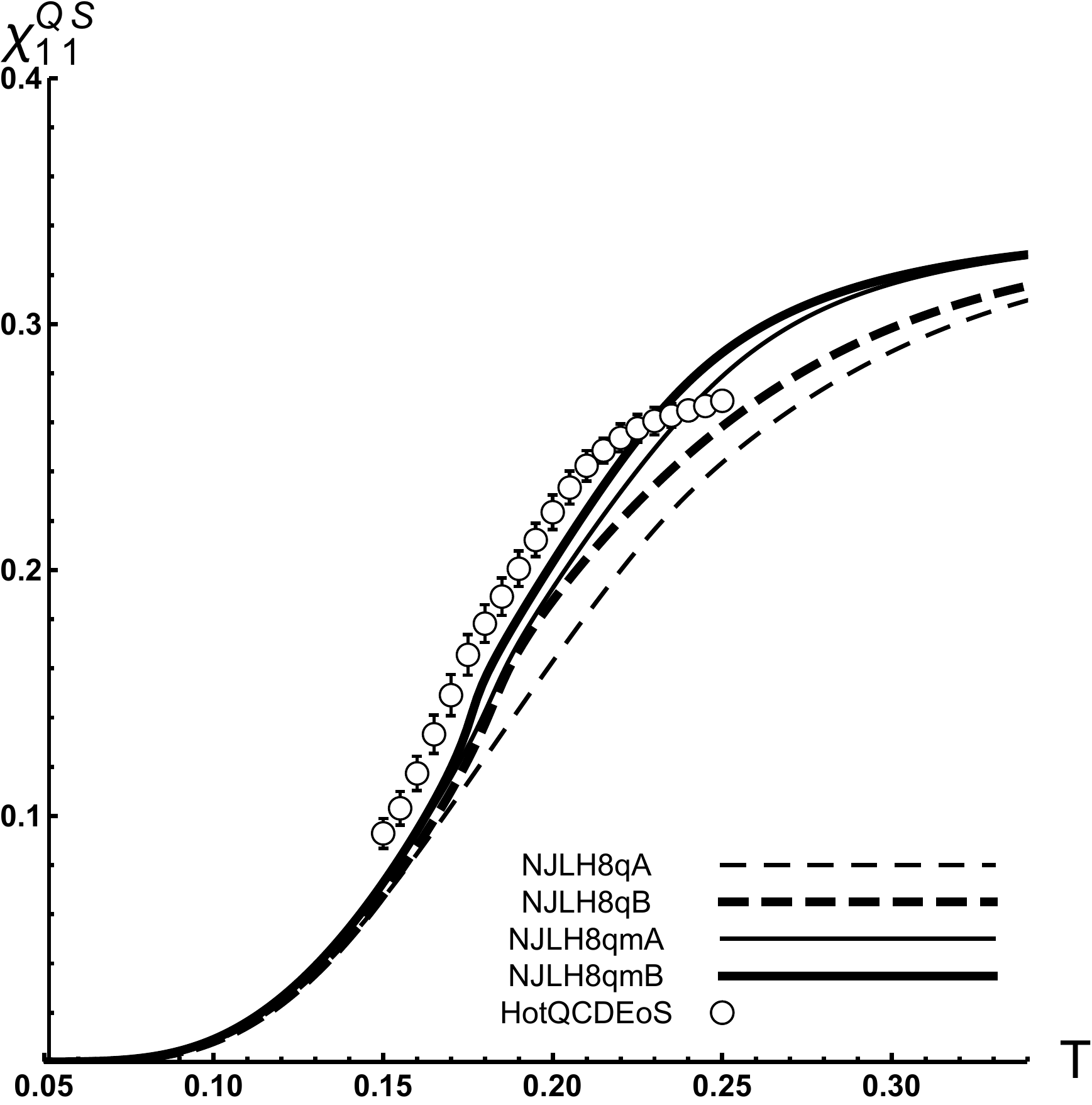}}
\caption{
Polyakov loop extension using ${\cal U}_{II}$ in Eq. \ref{UPII} for the same observables as in \ref{Chi2BQPELAPVLogTc200},\ref{Chi2BSPELAPVLogTc200},\ref{Chi2QSPELAPVLogTc200}. Notation as before.}
\label{CorrelationsPELAPVExpKLogBhatTc175}
\end{figure*}

Turning to the fluctuations and correlations of the different charge numbers $N_B,N_Q,N_S$ one observes that all observables which had a good slope in the NJL model, i.e. $\chi_2^B,\chi_2^S,\chi_{11}^{BS},\chi_{11}^{QS}$ for the sets with ESB interactions, get shifted to higher temperatures and agree fairly well with the lQCD data for both Polyakov loop potential implementations, see Figs.  \ref{Chi2BPELAPVLogTc200}, \ref{Chi2SPELAPVLogTc200}, \ref{Chi2BSPELAPVLogTc200}
\ref{Chi2QSPELAPVLogTc200} for the case with ${\cal U}_I$ and Figs. \ref{Chi2BPELAPVExpKLogBhatTc175},\ref{Chi2SPELAPVExpKLogBhatTc175},\ref{Chi2BSPELAPVExpKLogBhatTc175},\ref{Chi2QSPELAPVExpKLogBhatTc175} for the case  ${\cal U}_{II}$ respectively. Both potentials improve on the height of the peak of the correlator $\chi_{11}^{QS}$, bringing it closer to the lQCD result, compare Fig. \ref{Chi2BQELAPV} without Polyakov loop to Figs. \ref{Chi2BQPELAPVLogTc200} and \ref{Chi2BQPELAPVExpKLogBhatTc175} with ${\cal U}_I$ and  ${\cal U}_{II}$ respectively.

Unfortunately the slope of the correlator $\chi_2^Q$ is not improved with either potentials, see Figs.   \ref{Chi2QPELAPVLogTc200} and \ref{Chi2QPELAPVExpKLogBhatTc175}.

\begin{figure*}
\center
\subfigure[]{\includegraphics[width=0.32\textwidth]{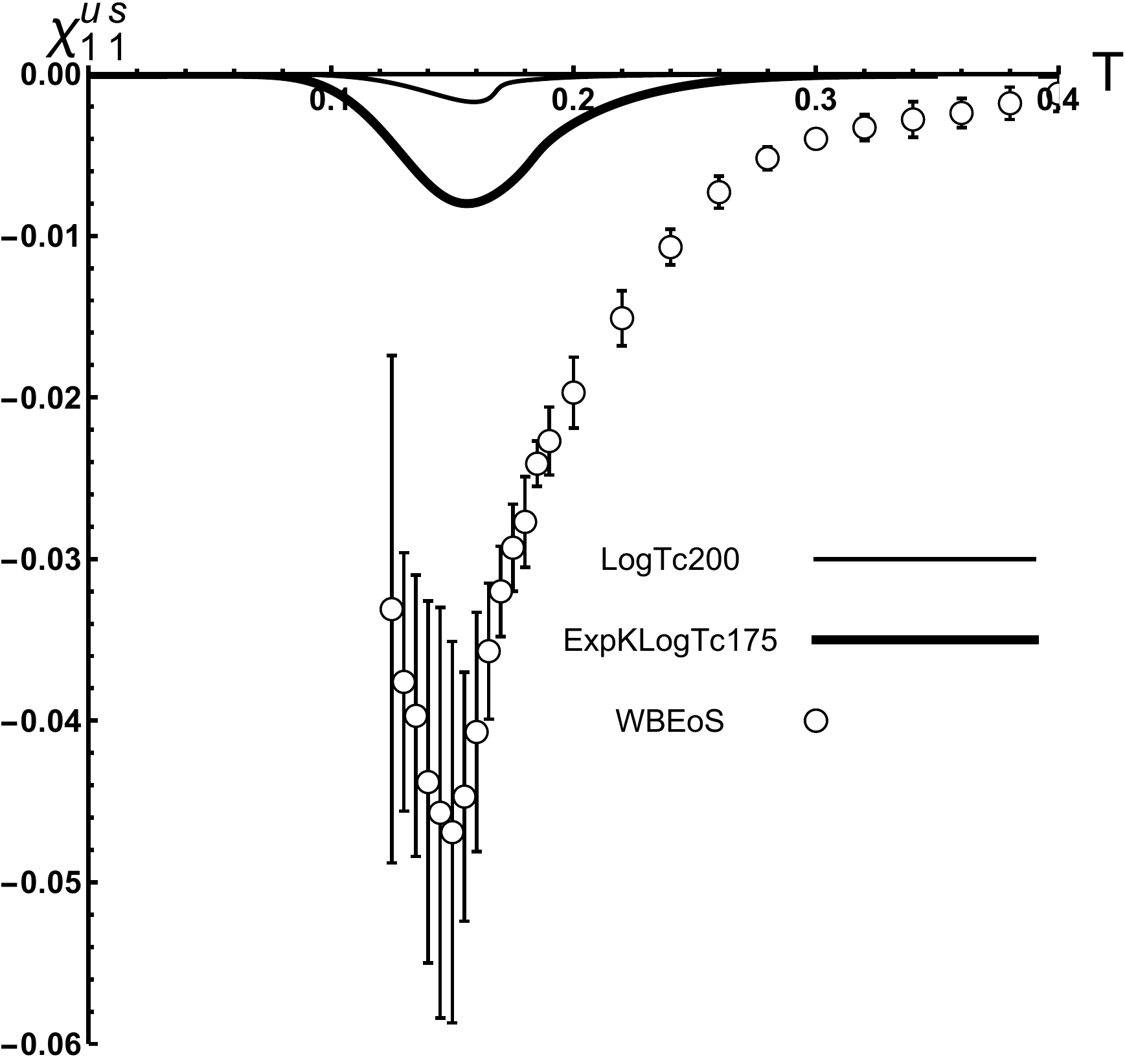}}
\caption{
Correlation of the up and strange charges ($\chi^{u\;s}_{1\;1}$) compared to continuum extrapolated lQCD results taken from \cite{Borsanyi:2011sw} (labeled as WBEoS). As this quantity is almost insensitive to the parametrization of the quark interactions we chose to display the effect of the choice of Polyakov potential (for the fermionic part we chose set NJLH8qB). Without the inclusion of Polyakov loop this quantity vanishes. 
}
\label{grafbwChi2us}
\end{figure*}

Finally we show in Fig. \ref{grafbwChi2us} that the coupling of the quark  and gluonic degrees of freedom leads to a non-vanishing correlation between the light and strange quark numbers, albeit smaller than in lQCD, with the ${\cal U}_{II}$ potential yielding a larger fraction. We also remark that this quantity is not sensitive to the details of the parametrizations in the quark sector.  
 
%%%%%%%%%%%%%%%%%%%%%%%%%%%%%%%%%%%%%%%%

\section{Conclusions}
\label{Conclusions}
We have used the three flavor NJL Lagrangian that has been enlarged in recent years to accommodate systematically current quark mass effects at NLO in the large $N_c$ counting scheme to address several thermodynamic observables. These explicit symmetry breaking (ESB) interaction terms are of the same order as the 't Hooft $U_A(1)$ breaking anomalous contribution and the previously  introduced symmetry preserving eight quark interactions. It has been shown that the ESB terms play a very important role in the description of accurate characteristics of pseudoscalar and scalar mesons. This opened for the first time the possibility to study the model phase diagram of QCD with a set of parameters which reproduces the empirical spectra, together with current quark masses that fit the actual PDG values, allowing to narrow down the uncertainties related to the model parameters.

While the model reaches systematically the Stefan-Boltzmann limit too fast as compared to the lattice results and is systematically shifted to lower temperatures as compared to lQCD, there are some relevant features which are reproduced. We highlight the main results:
\begin{enumerate}[(i)]
\item The ESB terms together with the realistic spectra select a region of parameters with strong OZI violating $8q$ coupling $g_1$. We recall that without the ESB terms there was an interval of values for this coupling, which in an interplay with the $4q$  G coupling, left the spectra unchanged except for the $\sigma(500)$ mass that got reduced for increasing $g_1$. The freedom in $g_1$ was accompanied by a sliding  CEP position in the model QCD phase diagram.   
\item In the strong $g_1$ coupling regime enforced by the ESB terms the velocity of sound displays a soft point as predicted by relativistic heavy ion models and lQCD. This relative minimum is absent in the NJL model which contemplates only the $4q$ and 't Hooft interactions, or weak $g_1$ couplings.
\item The trace of the energy momentum tensor displays a peak with height close to the lattice results; the slope of this quantity gets improved as compared to the model without the ESB interactions. However, although the strong $g_1$ coupling regime describes overall better slopes, it leads to a visible peak in the  transition regime for the related quantity $C_V$, which is not favored by lQCD data.
\item The slopes of the susceptibilities $\chi_2^{B}$, $\chi_2^{Q}$, $\chi_2^{S}$ of the conserved baryonic, electric and strange charges are sensitive to the weighting factors of the quark number susceptibilities $\chi_2^{u}$, $\chi_2^{d}$, $\chi_2^{s}$ that enter in their definition. We find that the slopes for $\chi_2^B$ and $\chi_2^S$,  as well as for the correlation involving these two charges, $\chi_{11}^{BS}$, get substantially improved, while it is too steep for $\chi_2^Q$. The observable $\chi_2^S$ is a clean probe for the slope of the strange quark susceptibility $\chi_2^s$, which agrees well with the corresponding lQCD slope.
\item Finally, by coupling the quark to the gluonic degrees of freedom via the Polyakov loop we observe that the temperature gap between the NJL  and the lQCD curves disappears practically and the overall characteristics of the lQCD data is rather well reproduced. For the trace of the energy momentum tensor the Polyakov loop potential ${\cal U}_{II}$ is better suited to describe the lQCD data than the potential  ${\cal U}_{I}$, within our model calculations.
\end{enumerate}

\section*{Acknowledgments}
Research supported by CFisUC and Fundac\~ao para a Ci\^encia e Tecnologia 
through the project UID/FIS/04564/2016,  and grants SFRH/BD/110315/2015,  SFRH/BPD/110421/2015.
We acknowledge networking support by the COST Action CA16201.

\bibliography{ThermoPropertiesNJLH8qm}
\end{document}